\documentclass[draft]{agujournal2019}
\usepackage{url} 
\usepackage{lineno}
\usepackage{lmodern}
\usepackage[finalnew]{trackchanges} 
\usepackage{soul}

\usepackage{adjustbox}
\usepackage{amsmath}
\newcommand{\tiunit}{TIU}

\draftfalse

\journalname{JGR: Planets}

\begin{document}

%
%


\title{Interactions of sublimated frost with volcanic plumes: Modelling Io's SO$_2$ atmosphere using the DSMC method}

%
%




\authors{Leander Schlarmann [1], Audrey Vorburger [1], Tim Mosimann [1], Nicolas~Thomas~[1], Peter Wurz [1]}

\affiliation{1}{Space Research \& Planetary Sciences, Physics Institute, University of Bern, 3012 Bern, Switzerland.}





\correspondingauthor{Leander Schlarmann}{leander.schlarmann@unibe.ch}



\begin{keypoints}
\item We conduct a literature review of thermophysical parameters for Io's surface frost to enhance a model of Io's sublimated SO$_2$ atmosphere.
\item Io’s eclipse may explain reduced column densities on the sub-Jovian hemisphere, while minor species 
could diminish sublimation-driven winds.
\item Volcanic plumes significantly alter Io’s atmosphere by displacing material, with the effect particularly pronounced on the dayside.
\end{keypoints}

%
%

%
%


\begin{abstract}
Io's tenuous atmosphere consists primarily of sulphur dioxide (SO$_2$) with observed column densities of approximately $10^{16}{-}10^{17}$~cm$^{-2}$. However, it remains uncertain whether the sublimation of SO$_2$ surface frost or volcanic outgassing is the primary source of the SO$_2$ atmosphere. In this study, we produce a 2D model of Io's SO$_2$ atmosphere using the Direct Simulation Monte Carlo (DSMC) method. For this purpose, we conduct a literature review on thermophysical parameters of Io's SO$_2$ surface frost to refine the surface temperature model in accordance with the most recent observations, which enables accurate modelling of Io's sublimated atmosphere. We find that the thermal conductivity, which shifts the peak temperature and SO$_2$ column density 
away from the subsolar point, has a pronounced effect on sublimation-driven winds and interactions with volcanic plumes. Furthermore, a background atmosphere {could} reduce sublimation-driven winds in Io's atmosphere, while the SO$_2$ column and number densities are not substantially altered. 
Moreover, we study the influence of the eclipse, when Io passes through Jupiter's shadow, finding that it reduces the average column density on the sub-Jovian hemisphere by a factor of ${\sim}5.5$ in relation to the anti-Jovian hemisphere. 
We also investigate the interaction of the sublimated atmosphere with a medium-sized plume at various locations relative to the subsolar point. We find a strong influence, especially on the dayside, where the atmosphere is enhanced with material being displaced by the plume, most pronounced when the plume is positioned near the point of maximum sublimation in the early afternoon. 
\end{abstract}

\section*{Plain Language Summary}
Io, the innermost of Jupiter’s large Galilean moons, is the volcanically most active body in our Solar System. Sulphur dioxide (SO$_2$) gas dominates Io's atmosphere, which may come from the sublimation of surface frost or from volcanic eruptions. In this study, we investigate the interactions of volcanic plumes with the sublimated atmosphere. Furthermore, we study how the atmosphere responds when Io moves into Jupiter’s shadow during an eclipse and how a background atmosphere of molecular oxygen (O$_2$) influences the sublimation-driven winds.

%
%

\section{Introduction}

Jupiter's satellite Io, the innermost of the Galilean moons, is subject to extreme tidal forces, making it the most volcanically active body in the Solar System. The satellite is comparable in size to Earth's Moon and is tidally locked to Jupiter. The first indication of Io’s active volcanism was the discovery of volcanic plumes in \textit{Voyager 1}  images \cite{morabito1979}. SO$_2$ was found to be a major atmospheric component \cite{hanel1979}. Before the Jupiter encounter, \citeA{peale1979} predicted that the orbital eccentricity and tidal heating generated by the Laplace resonance with Europa and Ganymede would lead to a largely molten interior for Io and speculated about active surface volcanism. Since the \textit{Voyager 1} observations, Io’s plumes have been studied with a variety of observational techniques, across different wavelengths, using ground-based and Earth-orbiting telescopes as well as space probes (e.g., \textit{Galileo}, \textit{Cassini}, \textit{New Horizons}).
Although volcanism is the ultimate source of Io's sulphur dioxide (SO$_2$) atmosphere, it remains uncertain whether volcanic eruptions or the sublimation of SO$_2$ frost is the dominant release process of atmospheric SO$_2$. 
Evidence for an atmosphere mainly supported by sublimation includes atmospheric changes during eclipse \cite<e.g.,>{clarke1994, geissler2004, saur2004, tsang2016, depater2020} and smooth variations in atmospheric abundance with latitude \cite<e.g., >{jessup2004}. Conversely, the lack of correlation between inferred column density and diurnal variation \cite<e.g.,>{strobel2001, jessup2015} could serve as evidence for a volcanic origin. 
{Moreover,} seasonal frost temperature variation over a Jovian year (11.86 Earth years) would be expected to influence the column density of the sublimated atmosphere, since Io receives over ${\sim}20$\% more solar flux at perihelion than at aphelion, given the strong temperature dependence of vapour pressure \cite{wagman1979, spencer2005}. {Observations by \citeA{tsang2012, tsang2013} and  \citeA{giles2024} of seasonal variations in the SO$_2$ column density point towards Io’s atmosphere being driven in a large part by sublimation.} {More details on Io’s atmosphere and plumes can be found in review papers by \citeA{lellouch2007} and \citeA{depater2021, depater2023}}. 

In this study, we model different source processes of Io’s SO$_2$ atmosphere using the Direct Simulation Monte Carlo (DSMC) method. We first present a literature study of thermophysical parameters of SO$_2$ frost on Io's surface to find the most realistic surface temperature profile, which we then use to model Io's sublimated SO$_2$ atmosphere. Furthermore, we investigate the influence of different processes on the sublimated SO$_2$ atmosphere. This includes the effects of the Jovian eclipse (Io passing through Jupiter's shadow) on the sub-Jovian hemisphere, the inclusion of a minor species (molecular oxygen, O$_2$) as a background gas, as well as the interactions of volcanic plumes with the sublimated atmosphere.

\subsection{Observations}

Several observations established that Io's atmosphere is spatially inhomogeneous, with SO$_2$ column densities on the order of $10^{16}{-}10^{17}$~cm$^{-2}$. Atmospheric species have been observed remotely at several wavelengths with ground- and space-based telescopes, providing estimates of column densities on the dayside.   
Ultraviolet (UV) observations were performed with different instruments on the Hubble Space Telescope \cite<HST;>{ballester1994, clarke1994, trafton1996, spencer1997, mcgrath2000, spencer2000, jessup2004, jessup2007} and include Lyman-$\alpha$ (Ly-$\alpha$) observations by \citeA{feldman2000, strobel2001, feaga2009}; and \citeA{giono2021} with the Space Telescope Imaging Spectrograph (STIS). 
In the infrared wavelength range, observations were performed with the NASA Infrared Telescope Facility \cite<IRTF;>{spencer2005, tsang2012, giles2024} and the ESO Very Large Telescope \cite<VLT;>{lellouch2015}.
Observations at millimetre (mm) wavelengths were performed with the 30 m IRAM radio telescope \cite{lellouch1990, lellouch1992, lellouch1996, lellouch2003, roth2020}, the Submillimeter Array \cite<SMA;>{moullet2010}, the Atacama Pathfinder EXperiment \cite<APEX;>{moullet2013}, and the Atacama Large Millimeter/submillimeter Array \cite<ALMA;>{depater2020, dekleer2024}. 

\begin{figure*} [t]
    \centering
    \includegraphics[width=0.99\textwidth]{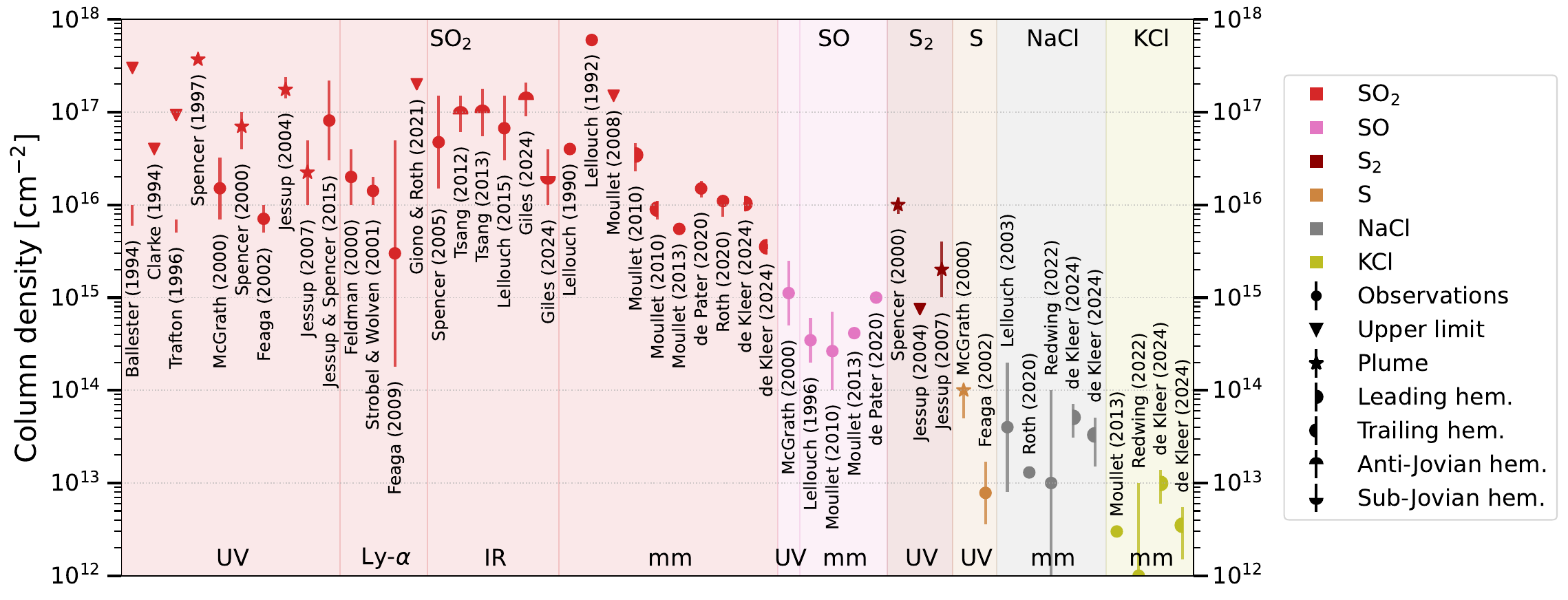}
    \caption{Column density of SO$_2$, SO, S$_2$, S, NaCl, and KCl derived from observations for Io. The observations are presented in chronological order and grouped according to their wavelength ranges (UV, Lyman-$\alpha$, IR, mm). Furthermore, the observed hemisphere is indicated, along with whether the observation targets a plume or reports only an upper limit. The vertical bars specify either the range or uncertainty of observed column densities. In Table~\ref{tab:observations}, the column densities and more detailed information on instruments and the wavelength range are given.}
    \label{Fig:columndensities}
\end{figure*}

Figure~\ref{Fig:columndensities} shows an overview of the column densities of different atmospheric species that have been observed in Io's atmosphere. Additional information on the respective observations can be found \ref{appendix:observations}.
The derived SO$_2$ column densities in Figure \ref{Fig:columndensities} cover a range of two decades, illustrating the variability of Io's atmosphere, which can be partially attributed to its volcanic activity \cite{spencer1997, spencer2000}. {SO$_2$ maps from ALMA observations by \citeA{depater2020} and \citeA{dekleer2024} clearly show the effect of volcanic activity.} 
{Inconsistencies in the observed SO$_2$ column densities also point to limitations of the different observation techniques}. Furthermore, asymmetries between the leading and trailing hemispheres \cite{moullet2010, dekleer2024} and the anti-Jovian and sub-Jovian hemispheres \cite{giles2024} were reported. 
Observed atmospheric species, apart from SO$_2$, include sulphur monoxide \cite<SO; >{lellouch1996}, molecular sulphur \cite<S$_2$; >{spencer2000}, atomic sulphur \cite<S; >{feaga2002}, atomic oxygen \cite<O; >{brown1981, wolven2001, roth2014}, sodium chloride \cite<NaCl; >{lellouch2003} and potassium chloride \cite<KCl; >{moullet2013}. NaCl and KCl are highly spatially confined, and the locations of volcanic plumes were identified as possible sources \cite{redwing2022, dekleer2024}.
Furthermore, material escaping from Io's atmosphere dominates the surrounding space environment \cite<see, e.g.,>{bagenal2020}, continuously supplying Io's plasma torus with fresh material as illustrated by \citeA{thomas2004}. Nevertheless, Figure~\ref{Fig:columndensities} demonstrates that the exact composition of Io’s atmosphere is still poorly constrained. 

\subsection{Modelling}

There have been various attempts to model Io's tenuous atmosphere. \citeA{ingersoll1985} modelled the flow of SO$_2$ from the dayside to the nightside using a hydrodynamic model
, which illustrated the possibility of huge pressure gradients at the terminator \cite<see also>{thomas1987}. 
Similarly, \citeA{moreno1991} modelled Io's sublimation and volcanic atmospheres, concluding that the nightside must be of volcanic origin and found that the sublimation region is located between the subsolar point and a solar zenith angle $\theta$ of $37^\circ$, while SO$_2$ condenses elsewhere. {\citeA{strobel1994} investigated the impact of solar heating,  plasma heating and Joule heating on the thermal structure of Io's atmosphere using a radiative-thermal conduction model. 
\citeA{wong1995} performed calculations for the effect of plasma bombardment on the sublimated atmosphere on the trailing hemisphere with a numerical fluid dynamic model, finding that plasma heating significantly inflates the exobase altitude. \citeA{wong1996_pc, wong1996_plasma} extended this work, also including photochemical products in the model and investigated the plasma effect on the surface. Moreover, \citeA{smyth2004} studied the impact of electron chemistry on Io's atmospheric composition and structure.}

\citeA{austin2000} first used the DSMC method to model Io's atmosphere, finding indications that a standing shock may form near the terminator. They also studied the effect of non-condensable gases by adding O$_2$ (and H$_2$S) at various gas pressures, finding a negligible influence on SO$_2$ deposition patterns. 
\citeA{zhang2003, zhang2004} performed axisymmetric DSMC simulations of plumes to demonstrate the formation of a canopy shock while studying vent conditions and matching observations. \citeA{moore2009} performed one-dimensional DSMC simulations to examine the effects of non-condensable species on a sublimating SO$_2$ atmosphere during eclipse. \citeA{walker2010} simulated Io’s sublimation-driven atmosphere in 3D, investigating the impact of plasma heating, planetary rotation, inhomogeneous surface frost, and molecular residence time of SO$_2$. \citeA{walker2012} performed a study of the thermophysical parameters and included effects of eclipse by Jupiter, finding that the atmospheric morphology of the sub-Jovian hemisphere is significantly affected by the "daily solar eclipse", causing a decrease in temperature (${\sim}5$~K) and column density (factor ${\sim}20$). \citeA{mcdoniel2015} performed 3D DSMC simulations to model the flow of SO$_2$ gas and silicate ash in Io's Pele plume, including the geometry of the lava lake, the umbrella-shaped canopy, the red deposition ring, and particle size. They demonstrated that the geometry of the lava lake can explain the shape of the plume's deposit pattern. \citeA{mcdoniel2017} investigated the interactions of giant plumes and the sublimated atmosphere. \citeA{mcdoniel2019} explored plume/plasma interactions with 3D DSMC simulations, finding that plasma inflates and heats plume canopies, being able to produce large diffusive neutral clouds. \citeA{ackley2021} and \citeA{hoey2016} performed DSMC simulations of the Tvashtar plume, investigating the influences of different vent aspect ratios.

In this study, we also use the DSMC method to model the interactions between volcanic plumes and the sublimated SO$_2$ atmosphere. However, while \citeA{mcdoniel2017} assumed a constant surface frost temperature (118~K) to model the sublimated SO$_2$ atmosphere for 2D and a simple exponential cosine surface temperature distribution of the form $\left[T_0 (\theta)=(118{-}70) \cos^{1/4}(\theta)  + 70\right]$ for 3D simulations, we use a more refined temperature model for the SO$_2$ frost on Io's surface. For this purpose, we conduct a literature {review} of thermophysical parameters for SO$_2$ frost, similar to \citeA{walker2012}, while incorporating new observations. Furthermore, we investigate the influences of an eclipse \cite<e.g.,>{walker2012} and a background gas \cite<e.g.,>{austin2000} on the improved model of the sublimated atmosphere.

\section{Methods}

The simulations in this study are performed using \texttt{ultraSPARTS} (\textbf{ultra}-fast \textbf{S}tatistical \textbf{PART}icle \textbf{S}imulation Package), a particle-based C\texttt{++} object-oriented parallel Direct Simulation Monte Carlo (DSMC) code by the Plasma Taiwan Innovation Corporation (Plasma T.I.) 
and is based on the DSMC code \cite<\texttt{PDSC++,}>{su2013, wu2003, wu2004, wu2005}. The code has been used in cometary research \cite{gerig2018, marschall2019,pinzon_rodriguez2021,mokhtari2025}.
The DSMC method, as described by \citeA{bird1994}, is physically accurate in all flow regimes; however, it becomes computationally expensive at high number densities \cite<e.g.,>{walker2010}. This is important for simulating Io's sublimated atmosphere, which is expected to be collisional near the subsolar point and volcanic plumes, but ballistic (free molecular flow) on the nightside, near the poles, and at high altitudes. The motion and collisions of \textit{computed} particles are simulated, representing a larger number of \textit{real} molecules and atoms. Our simulations were run with approximately $10^6$ computational particles, using particle weights between $10^{23}$ and $10^{25}$, which denotes the number of real particles represented per computational particle. For SO$_2${-}SO$_2$ and O$_2${-}O$_2$ collisions, we use reference diameters $d_\mathrm{SO_2}{=}7.16$~\AA\ and $d_\mathrm{O_2}{=4.07}$~\AA\ at 273~K, as well as viscosity indexes $\omega_\mathrm{SO_2}{=}1.05$ and $\omega_\mathrm{O_2}{=}0.77$ from \citeA{bird1994}. Furthermore, we include a Variable Soft Sphere \cite<VSS;>{koura1991} deflection parameter $\alpha=0.968$ for SO$_2${-}SO$_2$  collisions to make the scattering model more realistic. For collisions between molecules of different species (SO$_2$-O$_2$), the average diameter and viscosity index of the two involved species are used. 

We use \texttt{Fidelity Pointwise} as the mesh generator for our 2D grids, which enables us to optimise the grid's density, using tetrahedron-based unstructured grids. Our 2D grids have $3{-}9{\cdot}10^{5}$ cells with an average cell spacing of ${\sim}4$~km, using a transient adaptive sub-cell scheme implemented in \texttt{ultraSPARTS}. The simulations were run in parallel on 128 processors.
In the following parts, we describe how we model Io's sublimated atmosphere by conducting a literature study of thermophysical parameters for Io's surface to improve our thermal model. Furthermore, we show how we model volcanic plumes and discuss the caveats of our simulations in detail.

\subsection{Sublimation of SO$_2$ frost}

The temperature of the frost governs the saturation vapour pressure and, therefore, determines the occurrence of SO$_2$ sublimation or condensation \cite{doute2001}. The exponential relationship between temperature and SO$_2$ vapour pressure can lead to a significant change in SO$_2$ density for small temperature differences. In this study, we use the vapour pressure curves of \citeA{fray2009} to estimate the vapour pressure of SO$_2$. In \ref{appendix:vapour_pressure}, we study different measurements and curve fits for the vapour pressure of solid SO$_2$ found in the literature, to outline the justification for adopting the vapour pressure curves of \citeA{fray2009} in our simulations. We now describe the thermal model we used to estimate the surface temperature on Io, for which we conducted a literature study of thermophysical parameters. Furthermore, we describe the distribution of SO$_2$ frost on Io's surface, which is essential to modelling the sublimation.

\subsubsection{Thermal model} 

We use an adapted version of \texttt{THERMPROJRS}
, a 1-dimensional numerical thermal model which was developed by \citeA{spencer1989}.
The surface temperature is primarily governed by the balance between solar insolation, thermal radiation emitted from the surface, latent heat exchanges, and thermal conduction. \texttt{THERMPROJRS} enables us to include the ability of the surface to store heat in our calculations, which is characterised by the thermal inertia $\Gamma$ \cite{spencer1992}. The thermal inertia can be calculated from the thermal conductivity $\kappa$, the density $\rho$, and the heat capacity $C_p$ with Eq.~\ref{eq:thermal_inertia}.
\begin{equation}
    \Gamma = \sqrt{\kappa\,\rho\,C_p} \quad\left[\text{J m$^{-2}$ K$^{-1}$ s$^{-1/2}$}\right] \label{eq:thermal_inertia}
\end{equation} 
The unit of thermal inertia is J m$^{-2}$ K$^{-1}$ s$^{-1/2}$, which we denote as \tiunit\ throughout this study. However, the thermal inertia of Io's SO$_2$ frost is not well constrained, as the thermal conductivity $\kappa$ of solid SO$_2$ is not well studied. Moreover, the porosity of SO$_2$ frost on Io's surface, which influences $\rho$ and $\kappa$, is unknown. \citeA{kieffer1982} give a value of $2000$~kg/m$^3$ for the density of solid SO$_2$.  
Furthermore, $C_p$ can be interpolated for different temperatures from the data of \citeA{giauque1938}. For our calculations, we use a constant heat capacity $C_p=588.4$~J~kg$^{-1}$~K$^{-1}$ from \citeA{strobel1994}, as we find that the influence of the heat capacity on the temperature profile is limited. However, there is a large uncertainty for the density and specific heat capacity of the material on Io’s surface.

Another important thermophysical parameter governing the surface temperature is the {bolometric} Bond albedo, defined as the ratio {of} reflected solar energy in all directions {to} the total incoming solar energy. 
\citeA{simonelli2001} derived a global average surface Bond albedo of $0.52{\pm}0.09$ over Io's surface from analysing images from \textit{Galileo}'s solid-state imaging (SSI) camera. Before that, \citeA{simonelli1984} computed a Bond albedo of $0.5{\pm}0.1$ from Voyager full-disk images of Io, while \citeA{simonelli1988} obtained a value of $0.56$ for a patch of ``typical Ionian surface material''. However, these values do not constrain the albedo of pure SO$_2$ frost. Comparison of maps of the SO$_2$ frost coverage from \citeA{doute2001} and the Bond albedo \citeA{simonelli2001} suggests that the Bond albedo of pure SO$_2$ frost is larger than 0.55.

Furthermore, the main effect of a bolometric emissivity $\epsilon$ smaller than unity is an increased temperature, while it does not change the relative temperature variation \cite{tsang2015}. An emissivity of ${\sim}1$ is often considered a reasonable assumption for Io \cite<e.g.,>{matson1981, morrison1980, sinton1988}. Surface emissivities of 0.95 \cite{strobel1994} and 0.9 \cite{kerton1996, tsang2015, depater2020, dott2025} have been used in different studies. However, the true emissivity of Io’s SO$_2$ frost is not well constrained.

The latent heat of sublimation $L_{\text{sub}}$ refers to the amount of energy required to change the phase of a substance directly from the solid to the gaseous state. We use SO$_2$ latent heat of sublimation of $420$~kJ~kg$^{-1}$ \cite{dundas2017, milazzo2001}, which results from a value computed from critical table data by \citeA{nash1980} for SO$_2$ at 120~K and $10^{-8}$~bar.
\citeA{walker2012} found that the value of $L_{\text{sub}}$ has a relatively minor impact on temperature, which we confirm through a parameter study in \ref{sec:Parameter_study}.

The mass loss rate $\frac{\partial m}{\partial t}$ into vacuum can be calculated from the vapour pressure $P_v$ with the Hertz-Knudsen equation (Eq.~\ref{eq:Hertz_Knudsen}), where $m$ is the molecular mass in [kg], with the assumption of unit sticking coefficient for SO$_2$ \cite{persad2016}: 
\begin{equation}
    \frac{\partial m}{\partial t} = \frac{P_v}{\sqrt{2\pi\,k_B\,T/m}} \quad\left[\frac{\text{kg}}{\text{m}^2 \text{ s}}\right] \label{eq:Hertz_Knudsen}
\end{equation}

In \ref{sec:Parameter_study}, we investigate the influence of the Bond albedo ($A_B$), the emissivity ($\epsilon$), the thermal inertia ($\Gamma$), and the latent heat of sublimation ($L_{\text{sub}}$) on the diurnal temperature profile through a parameter study. However, since the exact values for the thermophysical parameters are still uncertain for Io's surface, we conduct a literature study to identify the most realistic model for Io's SO$_2$ frost temperature. 

\subsubsection{Literature study of thermophysical parameters}
\label{sec:Literature_study}

\begin{table*}[t]
\centering
\caption{Thermophysical parameters (thermal inertia $\Gamma$, Bond albedo $A_B$) for Io’s surface from different literature sources for a homogeneous and inhomogeneous two-component (frost and non-frost) surface thermal model. The unit of thermal inertia [J~m$^{-2}$~K$^{-1}$~s$^{-1/2}$] is referred to as \tiunit.}
\begin{adjustbox}{max width=\textwidth}
\begin{tabular}{ccccccccc}
\hline
\hline
\multicolumn{2}{c}{SO$_2$ frost component}     & \multicolumn{2}{c}{non-frost component} & Method & Volcanic & $\epsilon$& Notes  & Source  \\
$\Gamma$ [\tiunit]    & A$_B$                  & $\Gamma$ [\tiunit] & A$_B$              &        & [$10^{16}$ cm$^{-2}$] & & (e.g., hemisphere) & \\
\hline
56.65                 & 0.4746                 & 5.17         & 0.103          & eclipse &                     & 1   &              &\citeA{sinton1988} \\
26-92                 & 0.75                   & --           & --             & --      &                     & 0.9 &              &\citeA{kerton1996} \\
70                    & 0.52                   & --           & --             & diurnal &                     & 1$^{a}$& homogeneous  &\citeA{rathbun2004}$^b$\\
1000                  & 0.7                    & 40           & 0.34           & diurnal &                     & 1$^{a}$& inhomogeneous&\citeA{rathbun2004}\\
$200\pm50$            & $0.55\pm 0.02$         & $20\pm10$    & $0.49\pm0.02$  & --      &                     & 1   &              &\citeA{walker2012}$^b$ \\
300                   & 0.5                    & --           & --             & --      &                     & 1   & "unphysical" &\citeA{walker2012} \\
150-1250              & 0.613-0.425            & --           & --             & seasonal&                     & 0.9 &              &\citeA{tsang2012}  \\
150/200               & 0.613/0.575            & --           & --             & seasonal&                     & 0.9 & preferred    &\citeA{tsang2012}$^b$ \\
100-800               & 0.65-0.49              & --           & --             & seasonal&                     & 0.9 &              &\citeA{tsang2013}  \\
350                   & 0.62                   & --           & --             & eclipse & 1.8                 & 0.9 &              &\citeA{tsang2015}  \\
1000                  & 0.55                   & --           & --             & eclipse &                     & 0.9 & pure subl.   &\citeA{tsang2015}$^b$ \\
320                   & 0.5                    & 50           & 0.5            & eclipse &                     & 1$^{a}$&              &\citeA{depater2020}$^b$ \\
230                   & 0.565                  & --           & --             & seasonal& 7.4                 & 0.9$^{a}$& anti-Jovian  &\citeA{giles2024}  \\
$250^{+100}_{-90}$    & $0.56^{+0.04}_{-0.11}$ & --           & --             & seasonal& $7.4^{+0.9}_{-1.1}$ & 0.9$^{a}$& anti-Jovian  &\citeA{giles2024}$^b$ \\
60                    & 0.690                  & --           & --             & seasonal& 0.6                 & 0.9$^{a}$& sub-Jovian   &\citeA{giles2024}  \\
$80^{+420}_{-20}$     & $0.66^{+0.02}_{-0.06}$ & --           & --             & seasonal& $0.6^{+0.2}_{-0.6}$ & 0.9$^{a}$& sub-Jovian   &\citeA{giles2024}  \\
272-333 (298)         & 0.62 (0.5)             & --           & --             & --       &                    & 0.9 &              &\citeA{dott2025}$^b$ \\
\hline
\hline
\multicolumn{9}{c}{Notes. $^{a}$assumed bolometric emissivity. $^{b}$thermophysical parameters used for Figure~\ref{Fig:comparison_temperature_profiles}. }\\
\end{tabular}
\end{adjustbox}
\label{tab:thermophysical_parameters}
\end{table*}

Many studies of thermophysical parameters for Io have been conducted. Here, we provide an overview of the different values used in previous studies. A broad range of thermophysical parameters can explain observations. In Table~\ref{tab:thermophysical_parameters}, an overview of the values by different studies is given. 

One method for determining the thermophysical parameters of Io's surface is to analyse observations taken during its eclipses by Jupiter. \citeA{sinton1988} fitted thermophysical parameters to observations of eclipses with the NASA Infrared Telescope Facility (IRTF). Their best fit included a bright, high-thermal-inertia, thin layer region ($\Gamma{=}56.65$~\tiunit; $A_{B}{=}0.4746$) associated with SO$_2$ frost and a dark, low-thermal-inertia, homogeneous region ($\Gamma{=}5.17$~\tiunit; $A_{B}{=}0.103$) associated with "yellowish material" assumed to be sulphur. Furthermore, \citeA{depater2020} observed Io going into and out of eclipse with ALMA. They assumed Io's crust to be composed of two layers: a thin, few millimetre thick, low-thermal-inertia  ($\Gamma{=}50$~\tiunit, $A_B{=}0.5$, $\epsilon{=}0.9$) layer, overlying a more compact layer of high-thermal-inertia (320~\tiunit, $A_B{=}0.5$, $\epsilon{=}0.78$) layer of rock and/or coarse-grained or sintered ice. Therefore, the thermal inertia may vary over depth scales, which could influence the sublimation and condensation rates. \citeA{tsang2015} reported a lack of post-eclipse atmospheric changes from HST/COS observations and stated that their egress data cannot be fit with low thermal inertia material (${<}500$~\tiunit).  The closest match to their data for a purely sublimation-supported atmosphere was achieved with a high thermal inertia ($\Gamma{=}1000$~\tiunit) and a moderate Bond albedo ($A_B{=}0.52{-}0.55$). A purely volcanic supported atmosphere or a combination of frost with $\Gamma{=}350$~\tiunit, $A_B{=}0.62$ and a volcanic contribution with an equatorial density of $1.8\cdot10^{16}$~cm$^{-2}$ could also reproduce their observations. {Note that these observations were binned to 10-minute resolution, while \citeA{depater2020} showed that the atmosphere reforms within 10 minutes.}

Another approach focuses on observations of diurnal temperature variations. \citeA{rathbun2004} estimated the thermal inertia and Bond albedo from diurnal temperature variations measured with \textit{Galileo}'s photopolarimeter-radiometer (PPR) for both a homogeneous surface ($\Gamma{=}70$~\tiunit; $A_{B}{=}0.52$) and a inhomogeneous surface with a bright, high thermal inertia, frost component ($\Gamma{=}1000$~\tiunit; $A_{B}{=}0.7$) and a dark, low thermal inertia non-frost component ($\Gamma{=}40$~\tiunit; $A_{B}{=}0.34$), connected with dark pyroclastic dust, with an areal coverage of 50\%, respectively.

Moreover, seasonal temperature variations can be observed over the course of the Jovian year. \citeA{tsang2012} found that measurements with TEXES (Texas Echelon Cross Echelle Spectrograph) at the NASA IRTF of variations in atmospheric density over almost a full Jovian year (2001--2010) are best fit by thermal inertias between $150{-}1250$~\tiunit\ and Bond albedos between 0.613{--}0.425. However, they stated that photometric evidence favours albedos $A_B{>}0.55$. Therefore, their most preferred scenarios include thermal inertias of $150$ or $200$~\tiunit\ and Bond albedos of $0.613$ or $0.575$, with a constant volcanic component of $6{\cdot}10^{16}$~cm$^{-2}$ or $5{\cdot}10^{16}$~cm$^{-2}$ that was added to the time-varying sublimation component. \citeA{tsang2013} refined the analysis using data from 2012 and 2013, to constrain the thermal inertia between $100{-}800$~\tiunit, the Bond albedos between $0.65-0.49$, and a volcanic component between $(0.55{-}0.75)\cdot10^{17}$~cm$^{-2}$. \citeA{giles2024} also included IRTF/TEXES observations up until 2023. They report a best-fit model with a Bond albedo of 0.565, a thermal inertia of 230~\tiunit, and a constant (volcanic) component of $0.74{\cdot}10^{17}$ cm$^{-2}$. They also provided $\Gamma{=}250^{+100}_{-90}$~\tiunit, $A_B{=}0.56^{+0.04}_{-0.03}$, with a volcanic component of $0.74^{+0.09}_{-0.11}{\cdot}10^{17}$~cm$^{-2}$ for the anti-Jovian hemisphere (110.4--116.5 K). For the sub-Jovian hemisphere, they found best-fit parameters from their seasonal model of $\Gamma{=}80^{+420}_{-20}$~\tiunit, $A_B{=}0.66^{+0.02}_{-0.06}$, with a volcanic component of $0.06^{+0.02}_{-0.06}{\cdot}10^{17}$~cm$^{-2}$. Furthermore, they used $\Gamma{=}60$~\tiunit, $A_B{=}0.690$, and a constant component of $6{\cdot}10^{15}$ cm$^{-2}$. 

\citeA{kerton1996} modeled Io's frozen SO$_2$ surface for a typical value of the albedo of SO$_2$ frost $A_B{=}0.75$. They used a specific heat capacity of $470{-}1050$~J~kg$^{-1}$~K$^{-1}$, a thermal conductivity of $0.005{-}0.01$~J~m$^{-1}$~s$^{-1}$~K$^{-1}$, and a density of $300{-}800$~kg/m$^3$ for the porous ice regolith. From these values, one can calculate a thermal inertia $\Gamma{=}26{-}92$~\tiunit.
\citeA{walker2012} performed a parametric study on the thermophysical parameters of Io's surface. Their best-fit parameters were $\Gamma{=}200\pm50$~\tiunit\ and  $A_B{=}0.55\pm0.02$ for the SO$_2$ frost surface, as well as $\Gamma{=}20 \pm 10$~\tiunit\  and $A_{B}{=}0.49{\pm}0.02$ for the non-frost surface. They suggested volcanically ejected pyroclastic dust or sulphur allotropes as the non-frost composition. In addition, \citeA{walker2012} adopted a minimum SO$_2$ frost albedo of $0.55$ based on correlations between regions of high mean albedo \cite{simonelli2001} and high frost fraction \cite{doute2001}. Their best (but unphysical) fit was $\Gamma{=}300$~\tiunit\ and $A_B{=}0.5$. \citeA{dott2025} developed a time-dependent temperature model, using a default value of $298$~\tiunit, and a surface albedo of 0.5 in IR (0.62 in visible).

\begin{figure} [t]
    \includegraphics[width=0.99\textwidth]{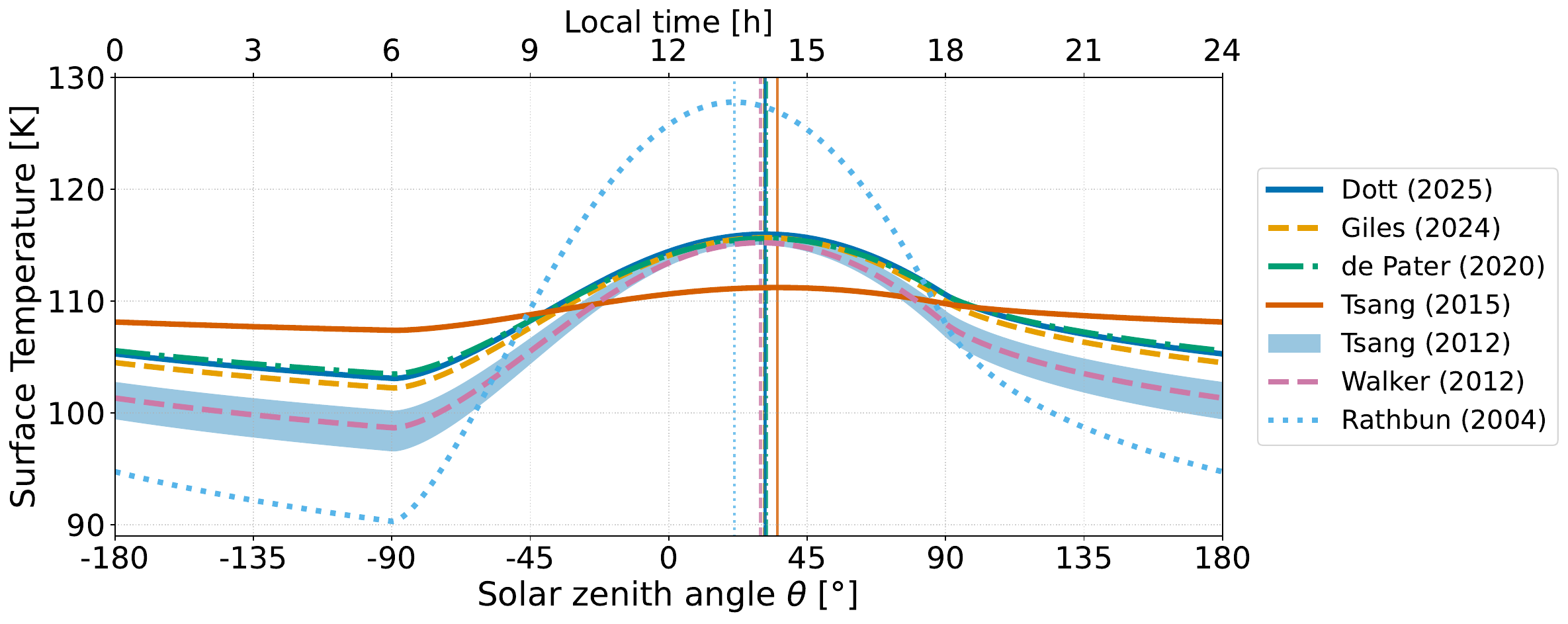}
    \caption{SO$_2$ frost temperature profiles of Io's surface with different thermophysical parameters from \citeA{walker2012}, \citeA{tsang2012, tsang2015}, \citeA{depater2020}, \citeA{giles2024}, and \citeA{dott2025}. {In addition, the temperature profile from the thermophysical properties of the homogeneous model by \citeA{rathbun2004} is shown, which also includes the non-frost component.} These temperature profiles were calculated using an adapted version of \texttt{THERMPROJRS}. The vertical lines refer to the shift of the maximum temperature from the subsolar point ($\theta{=}0^\circ$), which is specified in Table~\ref{tab:sub_atm_column_densities} together with the minimum and maximum temperature of the profiles.}
    \label{Fig:comparison_temperature_profiles}
\end{figure}

Overall, a wide range of thermophysical parameters for Io's SO$_2$ surface frost {are} found in the literature, with thermal inertias ranging from {56.65} to 1000~\tiunit, Bond albedos between 0.425 and 0.7{5} and emissivities between 0.9 and 1.
In Figure~\ref{Fig:comparison_temperature_profiles}, the temperature profiles resulting from the thermophysical parameters specifically for SO$_2$ frost in Table~\ref{tab:thermophysical_parameters} starting with \citeA{walker2012} are shown. 
In addition, we show the temperature profile of the homogeneous model by \citeA{rathbun2004}, with a low thermal inertia of 70~\tiunit. 
The thermophysical parameters of \citeA{rathbun2004} are based on thermal radiation measurements with \textit{Galileo} PPR, which are a superposition of surface regions covered with SO$_2$ frost and warm non-frost-covered areas. As we find that the latent heat of sublimation can have a strong influence on the resulting temperature profile for such low thermal inertias (see Figure~\ref{Fig:comparison_thermal_parameters}), we assumed a zero latent heat of sublimation for the temperature profile by \citeA{rathbun2004}, while using $L_{\text{sub}}{=}420$~kJ/kg \cite{milazzo2001} for the other temperature profiles. 

The resulting temperatures for the SO$_2$ frost are between 95~K and 116~K, with the maximum temperature shifted between 30$^\circ$ and $35^\circ$ from the subsolar point towards dusk. The determined maximum temperature of the SO$_2$ frost is lower than the upper limit of the temperature range of 90~K to 130~K of Io's surface measured with \textit{Galileo} PPR presented by \citeA{rathbun2004}, which also includes the radiation from hot spots and warm non-frost-covered areas.
The temperature profiles exhibit similar behaviour, with the minimum frost temperature varying between 97~K \cite[$\Gamma{=}150$~\tiunit]{tsang2012} and 107~K \cite{tsang2015}. The temperature range of \citeA{tsang2015} is generated with the parameters for a pure sublimation atmosphere, which has a high thermal inertia (1000~\tiunit) and, therefore, is much flatter than the others. The temperature profiles by \citeA{dott2025} and \citeA{depater2020} overlap almost perfectly, which is also expected as they use similar thermal inertias (298 and 320~\tiunit, respectively) and Bond albedoes (0.565 and 0.5, respectively). {It is physically expected that the thermal inertia varies with depth, being lowest at the surface and increasing with depth due to compaction \cite<e.g.,>{depater2020}. While diurnal/eclipse temperature variations are expected to be mainly caused by the top surface layers, seasonal variations probe deeper into the surface. Moreover, millimetre observations are expected to come from deeper layers, roughly a centimetre below Io’s surface, which is not probed by shorter wavelengths \cite{ferrari2018, depater2020}.} 


In our study, we used a thermal inertia $\Gamma{=}250$~\tiunit, a Bond albedo $A_B{=}0.565$, and an emissivity $\epsilon{=}0.9$ {to model Io's sublimated SO$_2$ atmosphere} based on the investigations of \citeA{giles2024}. 
{These thermophysical properties seem to be the least error-prone, as they result from observations of the seasonal variability over the course of nearly two Jovian years. Even though seasonal variations are expected to probe deeper layers than eclipse and diurnal variations, the resulting temperature curve in Figure~\ref{Fig:comparison_temperature_profiles} aligns well with eclipse observations \cite<e.g.,>{depater2020} and other models \cite<e.g.,>{dott2025}.} Furthermore, we used a latent heat of sublimation $L_{\text{sub}}{=}420$~kJ/kg from \citeA{milazzo2001} to model the temperature profile for Io's sublimated atmosphere.

\subsubsection{SO$_2$ surface frost distribution}

\citeA{doute2001} determined the SO$_2$ frost distribution and its grain size on Io's surface by analysis of hyperspectral image cubes from \textit{Galileo}'s near infrared mapping spectrometer (NIMS) covering three-fourths of the surface. 
They found that SO$_2$ frost covers 45\% of the imaged surface, which we incorporated in our simulations. This lies toward the higher end of results from \citeA{mcewen1988}, who analysed UV-VIS Voyager multispectral mosaics and found 30--50\% SO$_2$ coverage. Furthermore, \citeA{carlson1997} analysed a \textit{Galileo} NIMS spectro-image (anti-Jovian hemisphere) at different NIR wavelengths and found SO$_2$ almost everywhere, with spatially variable concentrations.
However, while \citeA{mcewen1988} and \citeA{carlson1997} found that SO$_2$ frost is located predominantly near the equator (but relatively scarce around Pele-type volcanoes and the polar regions), \citeA{doute2001} reported SO$_2$ to be concentrated within several large areas centred at mid-latitudes and correlated with the locations of the plumes. \citeA{doute2001} argued that the observed discrepancies were caused by the different detection methods and due to their sensitivity to grain size. \citeA{laver2008, laver2009} used ground-based telescopes (VLT/SINFONI and Keck/OSIRIS) to map the equivalent widths of two SO$_2$ frost absorption bands, showing a strongly equatorial distribution. 
{The coverage of the SO$_2$ surface frost map from \citeA{laver2008, laver2009} was later completed by \citeA{depater2020spat}, combining it with new Keck observations.}
\citeA{trumbo2022} used HST/STIS to map observed features of SO$_2$ in spatially resolved UV-visible spectra, finding extensive equatorial and low latitude deposits of SO$_2$. Nevertheless, for our simulations of Io's sublimated SO$_2$ atmosphere, we adopt the average frost coverage from \citeA{doute2001}{, by multiplying the sublimated flux with the average frost surface concentration of 45\%}.

\subsection{Volcanic plumes}

\citeA{strom1982} discuss plume observations by Voyager~1, where they document plume heights from $60{-}300$~km, which corresponds to ejection velocities of approximately $500{-}1000$~m/s. These ejection velocities were estimated from a ballistic model, where the particles follow parabolic trajectories, with $v{=}\sqrt{2\,g\,h_{\text{max}}}$, where $g$ denotes the gravitational acceleration at Io's surface 
and $h_{\text{max}}$ the measured plume altitude.
To model a medium-sized volcanic plume, we assume a circular vent with a radius of 8~km, which is similar to the vent diameters used earlier \cite{zhang2003, mcdoniel2017}. However, the vent geometry is often poorly constrained and can be asymmetric \cite<e.g.,>[used a rectangular virtual vent with different aspect ratios]{hoey2021, ackley2021}. Furthermore, we use a surface temperature of 300~K, a {vertical ejection} velocity of 500~m/s, and a surface density of $3\cdot10^{17}$~m$^{-3}$ for the particles ejected from the plume. {These vent temperature and velocity correspond to the input parameters used by \citeA{zhang2003}, based on results for Prometheus-type plumes by \citeA{mcewen1983}. Therefore, we model} a medium-sized plume, with plume particles reaching heights of approximately $150{-}200$~km.
Particles are created at the surface according to a half-Maxwellian velocity distribution with a mean speed of $v{=}\sqrt{\frac{8\,k_B\,T}{\pi\,m}}$. The surface density, temperature, and velocity are presumed to be uniform across the entire vent {in our simulations}. 

\subsection{Caveats}

Our simulations aim to provide a comprehensive overview of Io's SO$_2$ atmosphere, including the main physical and chemical processes. These are the sublimation of SO$_2$ frost as well as the outgassing of SO$_2$ in volcanic plumes. However, as with all simulations, simplifications are necessary to reduce the computational costs. In this section, we discuss these caveats of our simulation and try to assess the influence of the neglected processes.

First of all, Io's three-dimensionality is neglected in our simulations. Therefore, especially asymmetric features, such as the three-dimensional structure of some plumes, e.g. seen at Pele \cite{mcdoniel2015}, and surface features, cannot be displayed in our simulation. \citeA{schenk2001} identified 115 mountains on Io, with an average altitude of 6.3~km, and the highest peak is Boösaule Montes with 17.5~km altitude. Compared to the maximum dayside scale height, $H{=}\frac{k_B\,T_{\text{max}}}{m_{\text{SO}_2}\,g_{\text{Io}}}\simeq27$~km, calculated for the maximum temperature $T_{\text{max}}{=}115.7$~K of the thermal model by \citeA{giles2024}, with the Boltzmann constant $k_b$, the SO$_2$ mass $m_{\text{SO}_2}$, and Io's acceleration of gravity $g_{\text{Io}}$, these elevations can become relevant. However, in our model, we assume Io to be spherical, {whereas} including topographical features will create local variations in the surface temperature by shadowing and can influence the gas flow field \cite{klaiber2024}. 

Furthermore, we do not include the interactions of the Jovian magnetic field with Io's atmosphere (e.g., electron-neutral and ion-neutral interactions) in our simulations, which cause plasma bombardment \cite{wong1995}. It remains unclear if the atmosphere is thin enough for magnetospheric corotating ions to penetrate through the atmosphere to hit the surface and eject particles via sputtering \cite<e.g.,>{sieveka1985}. Sputtering is also neglected in our simulations, but could be a source {of} salts \cite<e.g., NaCl and KCl,>{roth2020}. \citeA{moore2012} state that sputtering could become a major source for the atmosphere when the surface temperature drops below 108~K, which is the case for the nightside of Io's atmosphere. Note that, for ion-induced sputtering, the majority of sputtered particles are atoms; thus, sputtering would predominantly contribute to S and O in the atmosphere.  
We also do not consider different escape processes in our simulations, as for heavy SO$_2$ molecules, Jeans escape is expected to be negligible.  Additionally, dissociation and ionisation due to the interaction with photons and electrons are not considered. These are expected to be the main production mechanisms for minor species (e.g., S, O, SO), which we do not model in our simulations. 

For simplification, Io's rotation is neglected in the simulations. Especially, the changing plasma flow direction relative to the subsolar point could be relevant for the interaction of the sublimated atmosphere with the plasma. 
The thermal radiation from Jupiter is not included in the model, which is expected to contribute ${\sim}0.55$~W/m$^2$ on the sub-Jovian point \cite{dott2025}. We also do not investigate the effects of the varying heliocentric distance (4.95–5.458 AU) on the surface temperature, which results in a change of the solar illumination of ${\sim}46$~W~m$^{-2}$ to ${\sim}56$~W~m$^{-2}$ and, therefore, the surface temperature is expected to vary over the Jovian year. \citeA{dott2025} estimated this seasonal surface temperature variation to be around $5$~K. Io’s spin axis is tilted by ${\sim}3^{\circ} $ with respect to Jupiter’s orbital plane. \citeA{dott2025} found that this inclination causes hemispheric seasons with northern summer in perihelion and northern winter in aphelion.

\section{Results and Discussion}

\subsection{Sublimated atmosphere}

\begin{table*}[t]
\centering
\caption{{Average} day- and nightside column densities ($\overline{N}_{\text{dayside}}/\overline{N}_{\text{nightside}}$) for SO$_2$ in Io’s atmosphere for the different thermal models from Figure~\ref{Fig:comparison_temperature_profiles}. Furthermore, the maximum dayside column densities ($N_{max}$), {referenced column density values ($N_{ref}$)}, the minimum and maximum temperature ($T_{min}$/$T_{max}$), as well as the shift of the maximum temperature from the subsolar point ($\Delta \theta$) are shown.}
\begin{adjustbox}{max width=\textwidth}
\begin{tabular}{cccccccc}
\hline
\hline
\multicolumn{4}{c}{Column density [cm$^{-2}$]} & \multicolumn{2}{c}{Temperature [K]} & $\Delta \theta$ [°] & Reference \\
$\overline{N}_{\text{dayside}}$ & $\overline{N}_{\text{nightside}}$ & $N_{\text{max}}$ & $N_{\text{ref}}$ & T$_{\text{min}}$ & T$_{\text{max}}$ & &\\
\hline
4.0$\cdot$10$^{16}$ & 3.2$\cdot$10$^{15}$ & 9.5$\cdot$10$^{16}$ & (3.7/8.5)$\cdot$10$^{16}$ ($^{a}$) & 103.1 & 116.0 & 31.3° &  \citeA{dott2025} \\
3.5$\cdot$10$^{16}$ & 2.4$\cdot$10$^{15}$ & 8.5$\cdot$10$^{16}$ & (4/21)$\cdot$10$^{16}$ ($^{a, b}$) & 102.2 & 115.7 & 31.0° &  \citeA{giles2024} \\
3.6$\cdot$10$^{16}$ & 3.4$\cdot$10$^{15}$ & 8.4$\cdot$10$^{16}$ & (1.5${\pm}$0.3)$\cdot$10$^{16}$ ($^c$) & 103.5 & 115.6 & 31.8° &  \citeA{depater2020} \\
1.3$\cdot$10$^{16}$ & 6.5$\cdot$10$^{15}$ & 1.9$\cdot$10$^{16}$ & 5$\cdot$10$^{16}$ ($^d$)         & 107.4 & 111.2 & 35.4° &  \citeA{tsang2015} \\
3.0$\cdot$10$^{16}$ & 0.9$\cdot$10$^{15}$ & 9.4$\cdot$10$^{16}$ & (6.1-15.1)$\cdot$10$^{16}$ ($^e$) & 102.2 & 116.0 & 31.2° &  \citeA{tsang2012} \\
2.8$\cdot$10$^{16}$ & 0.8$\cdot$10$^{15}$ & 7.4$\cdot$10$^{16}$ & (1.4/6.3)$\cdot$10$^{16}$ ($^{a}$) &  98.7 & 115.2 & 29.8° &  \citeA{walker2012} \\
\hline
\hline
\multicolumn{8}{c}{{Notes: ($^a$) sub-Jovian/anti-Jovian hemisphere, ($^b$) values for perihelion; aphelion: (1/9)$\cdot$10$^{16}$~cm$^{-2}$,}} \\
\multicolumn{8}{c}{{($^c$) typical value, ($^d$) equatorial, ($^e$) sub-solar value near aphelion/perihelion}}
\end{tabular}
\end{adjustbox}
\label{tab:sub_atm_column_densities}
\end{table*}

\begin{figure} [h]
    \includegraphics[width=0.99\textwidth]{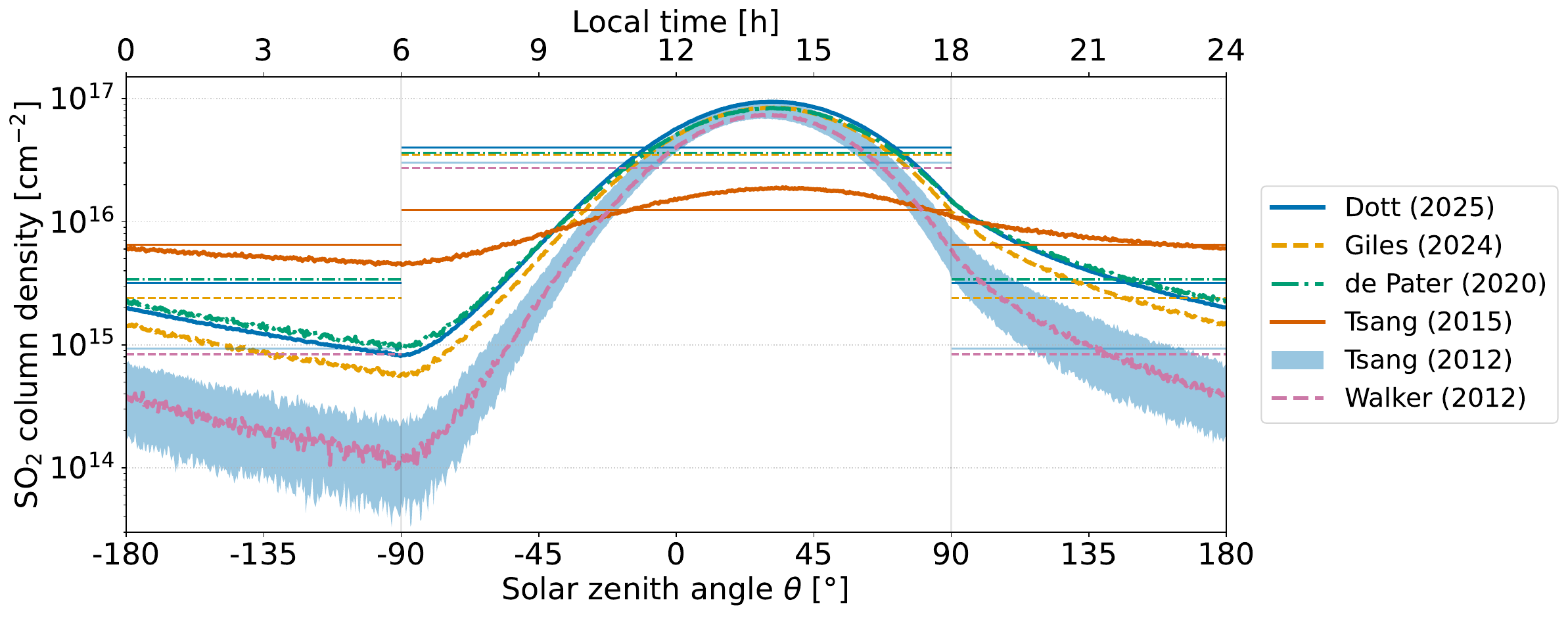}
    \caption{Column densities of SO$_2$ in the sublimated atmosphere of Io, calculated with the different thermophysical parameters specified in Table~\ref{tab:thermophysical_parameters}. Furthermore, the dayside and nightside average column densities from Table~\ref{tab:sub_atm_column_densities} are shown as horizontal lines.}
    \label{Fig:subl_atm}
\end{figure}

In this section, we investigate the influence of the {daily surface} temperature {variation} on the sublimation atmosphere. For this purpose, we compare the resulting column densities from our 2D DSMC simulations for the different thermophysical parameters used in the studies presented in Table~\ref{tab:thermophysical_parameters}. We focus on studies published within the past fifteen years, which provides a manageable yet representative dataset for our comparative analysis. In Figure~\ref{Fig:subl_atm}, we show the calculated column densities from thermophysical parameters by \citeA{walker2012, tsang2012,tsang2015, depater2020, giles2024, dott2025}. As one can see, the column density is predominantly governed by the temperature profiles, which are shown in Figure~\ref{Fig:comparison_temperature_profiles} for the different cases. In Table~\ref{tab:sub_atm_column_densities}, the average column densities on the dayside and nightside are shown. The results are comparable with the column densities of observations shown in Figure~\ref{Fig:columndensities}. We decided to use the thermal parameters of \citeA{giles2024} as our reference background atmosphere for the following study, as described in Section~\ref{sec:Literature_study}. {Our reference atmosphere} has an average dayside column density of $3.5{\cdot}10^{16}$~cm$^{-2}$, while the average nightside column density is 2.4$\cdot$10$^{15}$~cm$^{-2}$, which is about an order of magnitude lower. We now use this model to investigate the effect of different processes on the sublimated SO$_2$ atmosphere. This incorporates the effect of an eclipse, the inclusion of a minor species (O$_2$), as well as volcanic plumes.

\subsection{Eclipse}

Io's sub-Jovian hemisphere experiences solar eclipses during each orbit, due to the moon's tidally locked rotation around and proximity to Jupiter. The eclipse duration is approximately two hours \cite<7500 s;>{saur2004} with ${\sim}3$ min.\ of ingress and egress, which is significant compared with the orbital period of approximately 42.5 hours. Io’s passage into Jupiter’s shadow could also have implications for the satellite's atmosphere, since SO$_2$ frost would be expected to accumulate due to condensation. HST observations by \citeA{clarke1994} showed a decrease in brightness in UV lines by a factor 2--3 after Io passes into Jupiter's shadow. Observations of post-eclipse brightening by \citeA{binder1964} and \citeA{cruikshank1973} were the first indications of Io's atmosphere. They proposed that the atmosphere partially condenses on the surface in Jupiter's shadow. However, in subsequent observations, post-eclipse brightening was rarely seen \cite{bellucci2004, cruikshank2010}. Therefore, no clear explanation for the process, which was not always detected \cite<e.g., >{tsang2015}, has been provided. \citeA{depater2020} found an up to ${\sim}20\%$ atmospheric post-eclipse brightening in several SO$_2$ transitions 10 minutes after reemerging from eclipse into sunlight, which could be connected to the interaction with volcanic plumes.

\begin{figure*} [h]
    \includegraphics[width=0.99\textwidth]{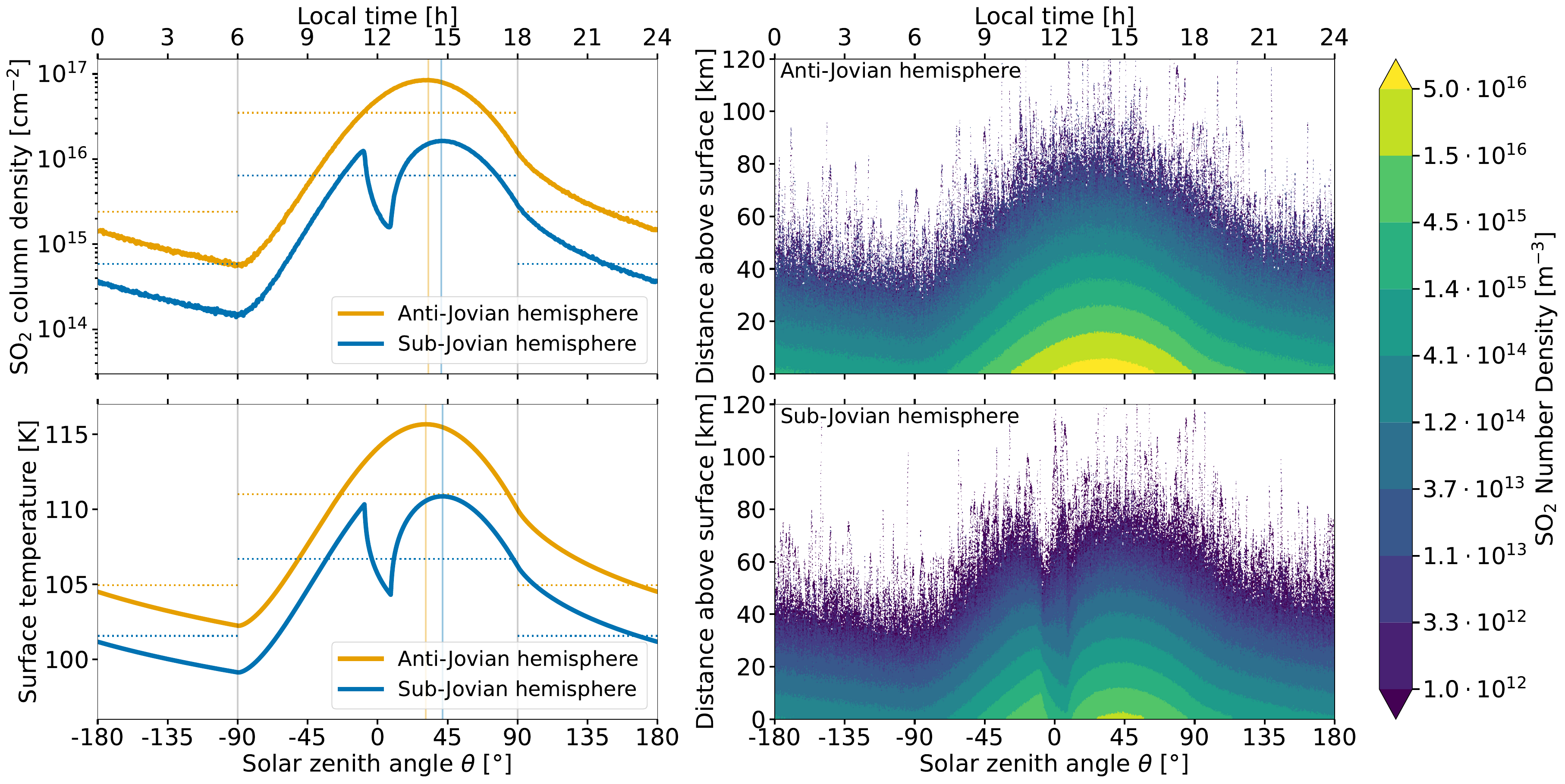}
    \caption{Comparison of the temperature profile and the SO$_2$ column and number densities on the anti-Jovian and the sub-Jovian hemisphere, whereas the sub-Jovian hemisphere experiences an eclipse with a duration of 2~h. Here, we illustrate the diurnal variations of these properties at the sub-Jovian and anti-Jovian points, resulting in an average dayside column density differing by a factor of ${\sim}5.5$.}
    \label{Fig:comparison_eclipse}
\end{figure*}

We model the eclipse of Io on the sub-Jovian hemisphere by setting the solar energy input to zero for two hours centred on the subsolar point, resulting in cooler average temperatures. The resulting temperature profile, therefore, represents the temperature change at the sub-Jovian point over the course of one orbit around Jupiter. In Figure~\ref{Fig:comparison_eclipse}, we compare the resulting SO$_2$ column density, number density, and surface temperature on the anti- and sub-Jovian hemisphere. The effects of the eclipse are evident, producing a dip in surface temperature near the subsolar point and consequently reducing both the number and column density of SO$_2$. The mean day- and night-side temperatures on the sub-Jovian hemisphere (106.7~K and 101.6~K) are significantly (about 3--4\%) lower than on the anti-Jovian hemisphere (111.0~K and 104.9~K). Furthermore, the offset of the peak temperature from the subsolar point is 10.7$^\circ$ higher on the sub-Jovian hemisphere (41.7°) than on the anti-Jovian hemisphere (31.0$^\circ$). The difference between the peak temperatures on the sub-Jovian and anti-Jovian hemispheres is 4.8~K, which is comparable to the results of simulations by \citeA{walker2012}, who find a temperature difference of 4.2~K. Furthermore, the total diurnal temperature variation is lowered by the eclipse in our simulation (11.8~K sub-Jovian; 13.5~K anti-Jovian). This is similar to results by \citeA{dott2025} (10.5~K; 12.3~K), \citeA{tsang2012} (10~K; 12~K), and \citeA{walker2010} (${\sim}14$~K; ${\sim}17$~K). \citeA{dott2025} find that the temperature decreases by up to 7.5~K within 10~minutes after the eclipse{, corresponding to an atmospheric collapse.} \citeA{tsang2016} observed the atmospheric collapse of Io’s atmosphere using Gemini/TEXES, reporting a decrease of the 19~$\mu m$ brightness surface temperature from 127~K to 105~K after 40 min in eclipse. \citeA{depater2020} report a weaker temperature decrease of ${\sim}$3~K in millimetre wavelengths from ALMA observations. 
In our simulation, the {SO$_2$} surface {frost} temperature decreases by approximately 6~K (from 110.3~K to 104.3~K) during the {2 hour} eclipse {duration, returning to its pre-eclipse temperature within 137 minutes.} 

Moreover, the average dayside column density is about $5.5$ times higher on the anti-Jovian ($35.0{\cdot}10^{15}$~cm$^{-2}$) than the sub-Jovian ($6.4{\cdot}10^{15}$~cm$^{-2}$) hemisphere {in our study}. On the nightside, the difference of the mean temperature is slightly smaller, with an increase by a factor of $4.1$ between the sub-Jovian ($5.9{\cdot}10^{14}$~cm$^{-2}$) and anti-Jovian ($24.1{\cdot}10^{14}$~cm$^{-2}$) hemisphere. The peak column densities also reflect this asymmetry, with $8.5{\cdot}10^{16}$~cm$^{-2}$ on the anti-Jovian hemisphere and $1.6{\cdot}10^{16}$~cm$^{-2}$ on the sub-Jovian hemisphere, reflecting an increase of a factor $5.3$. During the eclipse, the column densities in our simulation decrease by a factor of 7.8, from $12.5{\cdot}10^{15}$~cm$^{-2}$ to $1.6{\cdot}10^{15}$~cm$^{-2}$ over the duration of 2 hours. \citeA{tsang2016} {observe} that the SO$_2$ column density drops from $(2.0{-}2.5){\cdot}10^{16}$~cm$^{-2}$ to $0.5{\cdot}10^{16}$~cm$^{-2}$ during the eclipse, i.e., by a factor of $5\pm2$, which is slightly lower than our result. 
However, since we did not include ingress and egress times in our thermal model of the eclipse, a slight overestimation is also expected. 

An asymmetry between sub-Jovian and anti-Jovian hemispheres has been observed \cite<see, e.g.,>{dekleer2024, feaga2009, moullet2008, spencer2005}, where the sub-Jovian hemisphere consistently has lower column densities than the anti-Jovian hemisphere. Possible explanations could be the hemispherically asymmetry in the distribution of volcanic plumes, with more activity on the anti-Jovian hemisphere, or the inhomogeneous distribution of SO$_2$ frost. \citeA{walker2012} suggest the eclipse experienced by the sub-Jovian hemisphere as an additional explanation, which is supported by our results. 

\subsection{Sublimation-driven winds and minor species (O$_2$) }

On Io's nightside, there is the possibility of a localised collisional atmosphere due to volcanic activity \cite<e.g.,>{johnson1995}. Furthermore, non-condensible gases (O$_2$, SO) will migrate to Io's nightside. These gases are expected to form through photochemical reactions, electron impacts, or radiolysis and then accumulate on the nightside, creating a ``background'' nightside atmosphere. Additionally, SO could be released directly from volcanoes \cite{zolotov1998}. Keck observations by \citeA{depater2020spat} showed that the spatial SO distribution varies considerably across Io, suggesting that stealth volcanism 
causes the emissions. 
\citeA{kumar1982} and \citeA{ingersoll1993} already speculated that a collisional nightside O$_2$ atmosphere may accumulate on Io. 
\citeA{wong1996_pc} found that such a ``background'' atmosphere would reduce wind speeds and increase the atmospheric pressure. However, they state that the dayside SO$_2$ dynamics are not overwhelmed by the non-condensible gases in their model, with maximum column densities of non-condensible species about two orders of magnitude below the abundance of SO$_2$ on the dayside. \citeA{summers1996} state that photochemistry alone is unlikely to produce a collisional thick background atmosphere. 
\citeA{moses2002} included volcanic sources in their model, giving O$_2$ column densities between $10^{12}{-}10^{13}$~cm$^{-2}$ as an important SO$_2$ photolysis product. \citeA{smyth2004} find that the O$_2$ column density is enhanced by electron chemistry on the thicker dayside atmosphere, specifying column densities between $10^{14}{-}10^{15}$~cm$^{-2}$. Moreover, {\citeA{depater2023}} state that O$_2$ is predicted to be at the 1\% level of SO$_2${, based on \citeA{moses2002}}. 

\begin{figure*} [t]
    \includegraphics[width=0.99\textwidth]{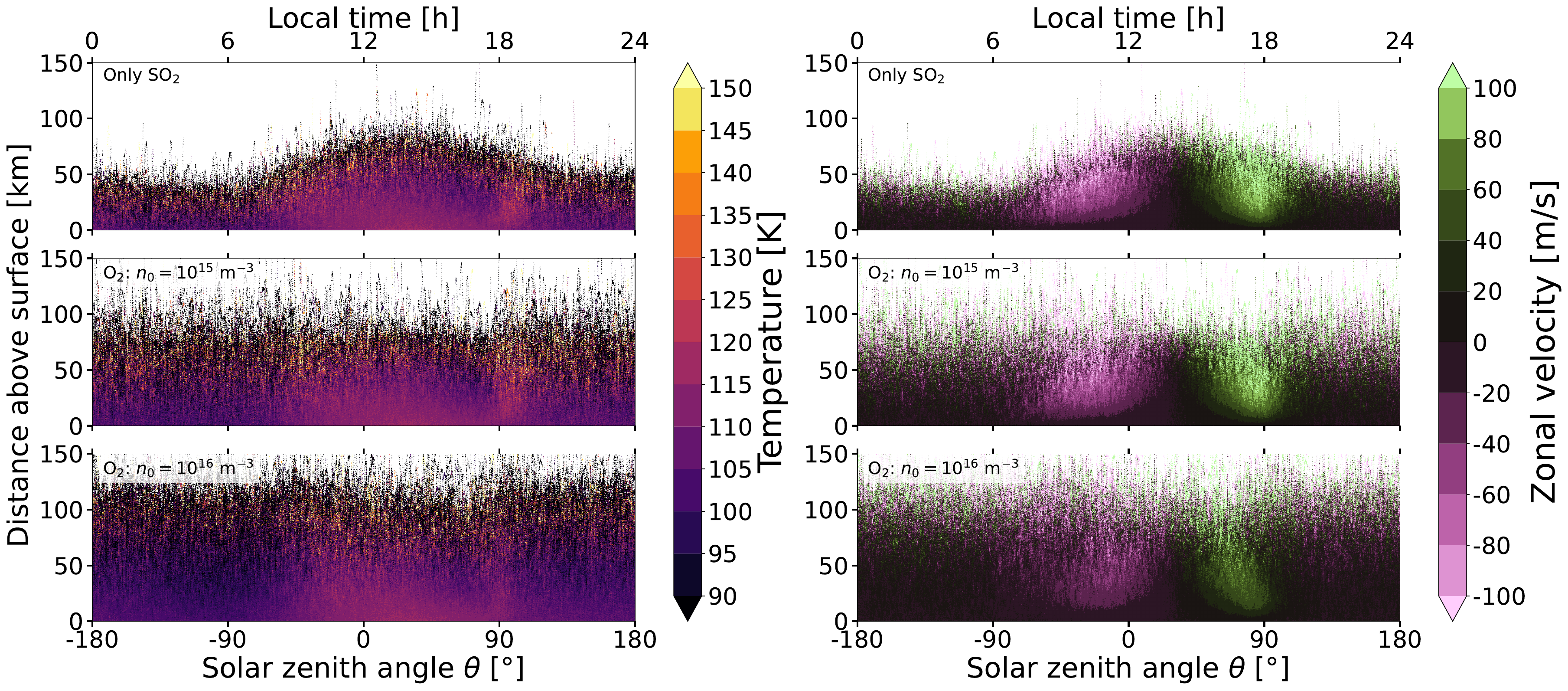}
    \caption{Comparison of the temperature, and {zonal} velocity for our reference SO$_2$ atmosphere with different surface densities of O$_2$ ($10^{15}$~m$^{-3}$, $10^{16}$~m$^{-3}$).{ In \ref{sec:Reference_atm}, the corresponding SO$_2$ and O$_2$ number densities are shown.}}
    \label{Fig:O2}
\end{figure*}

To investigate the influence of minor species, we model the SO$_2$ atmosphere, while varying the O$_2$ surface density from $10^{15}$~m$^{-3}$ to $10^{16}$~m$^{-3}$. Molecular oxygen has not been detected in Io's atmosphere so far, but O$_2$ is expected as a chemical product of SO$_2$ through different reactions, such as photolysis. In contrast to SO$_2$, the O$_2$ molecules do not freeze out when they return to the moon's surface, thus they could accumulate to significant levels in the atmosphere. In Figure~\ref{Fig:O2}, we compare the mean temperature and the zonal particle velocities in Io's atmosphere for different surface densities of O$_2$ as well as a pure sublimated SO$_2$ atmosphere without O$_2$. Additionally, the number densities of the modelled species (SO$_2$ and O$_2$) are shown in Figure \ref{Fig:reference_atm}. {The chosen surface densities are in the upper range, resulting in mean O$_2$ column densities of $1.3{\cdot}10^{15}$~cm$^{-2}$/$1.3{\cdot}10^{16}$~cm$^{-2}$ for O$_2$ surface densities of $10^{15}$~m$^{-3}$/$10^{16}$~m$^{-3}$.}
The zonal particle velocity refers to the component parallel to the surface in the equatorial direction, with eastward motion defined as positive. Therefore, it is equivalent to the equatorial zonal wind velocity.
We find that changing the surface density of O$_2$ in the investigated range does not alter the number and column density of sublimated SO$_2$. However, the mean temperature and zonal velocity decrease, especially at high altitudes, with increasing O$_2$ number density. This is expected, since the particles should follow the density gradient, meaning that the particles ‘flow’ towards lower densities. If the O$_2$ density increases, the overall density gradient is reduced, and these flows should become smaller or even stop.
These chosen surface densities are in the upper range of the expected column densities.

\begin{figure*} [t]
    \centering
    \includegraphics[width=0.99\textwidth]{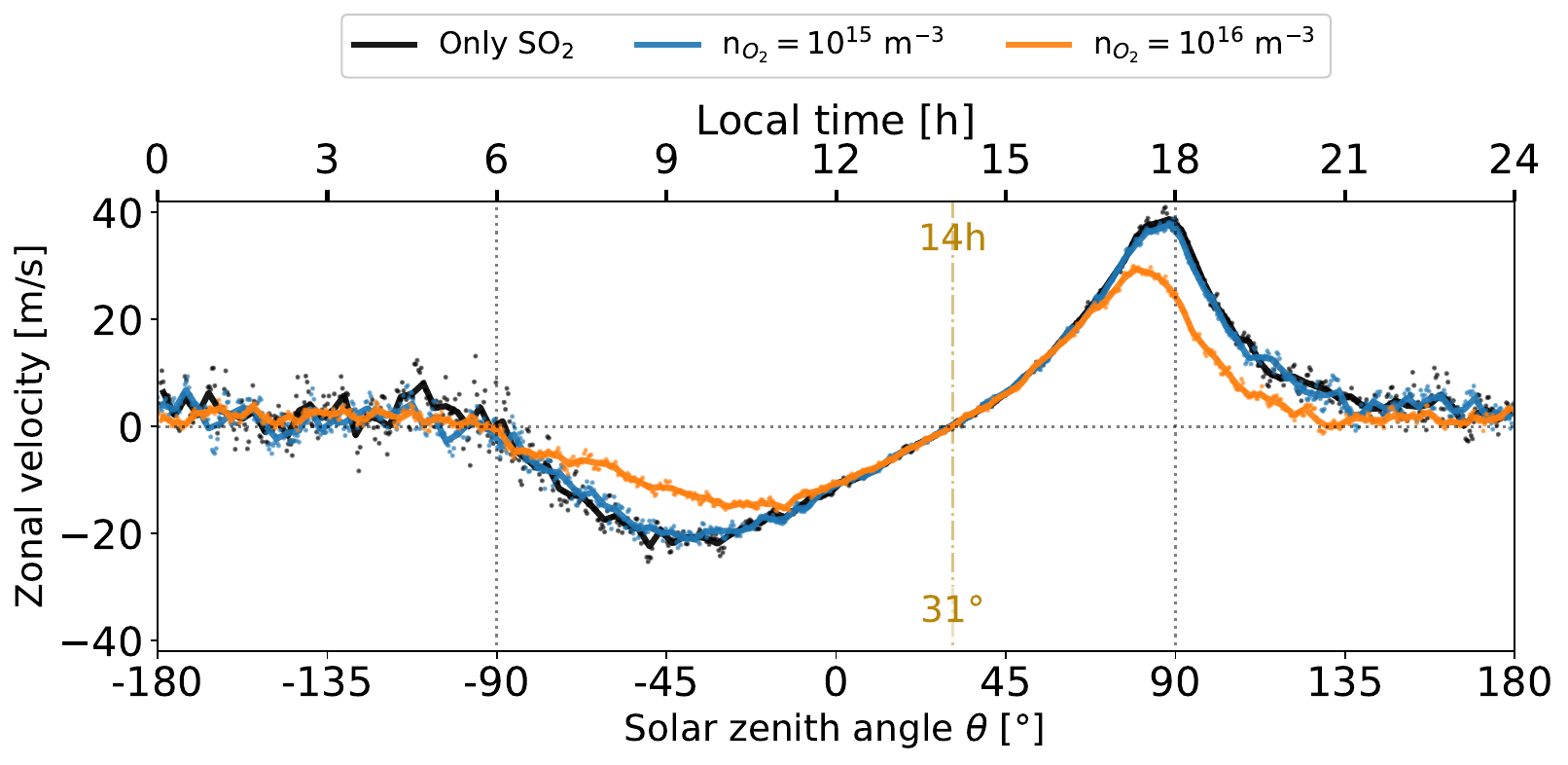}
    \caption{Comparison of the weighted mean of the {zonal} velocity $\langle v \rangle_{n}$ of SO$_2$ and O$_2$ molecules, with the number density used as the weight, for our reference atmosphere with different surface densities $n_{\text{O}_2}$ of O$_2$ ($10^{15}$~m$^{-3}$, and $10^{16}$~m$^{-3}$. The dash-dotted vertical line represents the point of maximal sublimation shifted 31° from the subsolar point. The sublimation-driven winds are directed away from this point, where $\langle v \rangle_{n}{=}0$~m/s, towards the respective terminator. The weighted mean of the {zonal} velocity for each simulated solar zenith angle (small dots) is averaged over 3° bins (solid lines).}
    \label{Fig:O2_zonal_velocity}
\end{figure*}

Furthermore, we calculate the weighted mean of the zonal particle velocity, weighted by number density.
In Figure~\ref{Fig:O2_zonal_velocity}, we show the average zonal velocity for all particles at different solar zenith angles for the different O$_2$ surface densities. Additionally, the zonal velocity of our reference atmosphere (with no O$_2$ molecules) is shown. One can see that the zonal velocity is directed away from the point of maximum sublimation (31° from the subsolar point), where it is zero, toward the respective terminator. This can be interpreted as sublimation-driven winds driven by the pressure and temperature gradient. However, the wind speed is reduced with increasing O$_2$ background gas, as the denser O$_2$ ``background" atmosphere dampens the velocity of SO$_2$ molecules. An asymmetry between the velocities of the sublimation-driven winds towards the nightside terminator (+90$^\circ$) and the morning terminator (--90$^\circ$). 
The range of the weighted mean wind velocities in Figure~\ref{Fig:O2_zonal_velocity} decreases from 66~m/s for the pure sublimated atmosphere, to 61~m/s ($n_{\text{O}_2}{=}10^{15}$~m$^{-3}$) {and} 46~m/s ($n_{\text{O}_2}{=}10^{16}$~m$^{-3}$), when O$_2$ is included as a background gas. {Therefore, a significant influence of the O$_2$ background atmosphere on the sublimation-driven winds is only seen at O$_2$ surface densities of $10^{16}$~m$^{-3}$, which exceed estimated values by \citeA{moses2002, smyth2004}}. 
However, we only show the sublimation-driven winds, while Io's equatorial rotation \cite<${\sim}75$~m/s; >{moullet2008}, and the rotation of the Jovian plasma torus, which overtakes Io at a relative velocity of $53-57$~km/s \cite{bagenal2020}, are neglected in our simulation.

Observations of winds in Io's atmosphere are sparse. \citeA{moullet2008} measured superrotating winds with limb-to-limb velocities of $330\pm100$~m/s in prograde direction on Io's leading hemisphere. \citeA{thelen2024} analysed Doppler shift maps from ALMA observations, finding wind line-of-sight velocities in the range of $25{-}50$~m/s in the SO$_2$ atmosphere within ${\sim}30^\circ$ of the subsolar point (receding). Furthermore, they find line-of-sight winds of up to $-100$~m/s and ${>}250$~m/s in approaching and receding direction.
\citeA{walker2012} presented integrated equatorial wind speeds of 150~m/s at the dusk terminator (prograde direction) and 100~m/s (retrograde direction) at the dawn terminator {in their simulations}, whereas Io’s equatorial rotation was also not accounted for. 
Additional observations of wind speeds in Io's atmosphere would be helpful to determine whether superrotating winds are present. 
\citeA{mcdonald2022} recently found similarities to dunes in linear features of the \textit{Galileo} probe, indicating aeolian sediment transport {and suggesting} that wind-blown transport via saltation could also alter Io's surface, {particularly through} interactions {between} lava and SO$_2$ frost. {However, the threshold friction speed required to move grains on Io was estimated by \citeA{bart2004} to be 20~km/s, which is orders of magnitude higher than the sublimated wind speeds found in this study.}

\subsection{Volcanic plumes}

\begin{figure*} [t!]
    \includegraphics[width=0.99\textwidth]{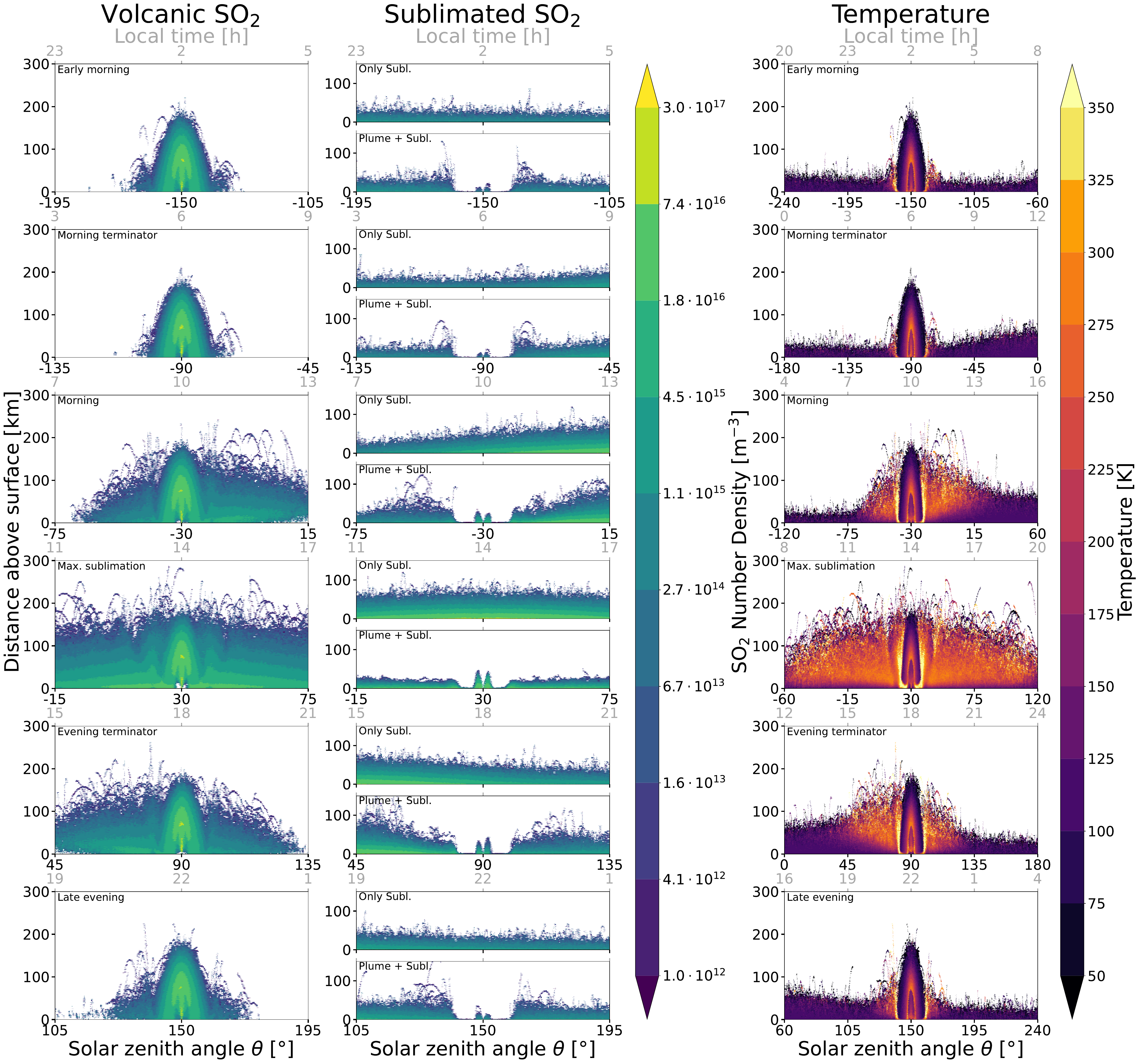}
    \caption{Interactions of Io's sublimated SO$_2$ atmosphere with a {medium-sized} plume for different angles relative to the subsolar point (0$^\circ$): --150$^\circ${/2h} (early morning), --90$^\circ${/6h} (morning terminator; min. sublimation), --30$^\circ${/10h} (morning), +30$^\circ${/14h} (afternoon; max. sublimation), +90$^\circ${/18h} (evening terminator), +150$^\circ${/22h} (late evening). A strong influence on the sublimated atmosphere is apparent on the dayside, where an enhanced SO$_2$ atmosphere can be observed. Stronger interactions occur at higher number densities of the sublimated atmosphere, which is reflected in the asymmetry in interactions when the plume is located at --30$^\circ$ and +90$^\circ$. When the plume location coincides with the point of maximum sublimation (+30$^\circ${/14h}), the entire hemisphere is affected by these interactions, which is reflected in the increased temperature. Contrary, on the nightside at +150$^\circ${/22h}, --150$^\circ${/2h} and --90$^\circ${/6h} interactions can only be observed in direct proximity of the plume.}
    \label{Fig:Plumes}
\end{figure*}

In Figure~\ref{Fig:Plumes}, we show the interaction of the sublimated atmosphere with a medium-sized plume. 
We define volcanic SO$_2$ separately from sublimated SO$_2$ by simulating them as different species with identical species parameters. This enables us to track particles originating from sublimation and volcanism separately. Therefore, the mass of the atmosphere that is added due to the volcanic eruption, defined as the volcanic component by \citeA{tsang2012} and \citeA{mcdoniel2017}, can be estimated. 
In Figure~\ref{Fig:Plumes}, we compare the number density and temperature of SO$_2$ molecules that were produced by sublimation or volcanic outgassing at the plume vent. For this purpose, we model the mid-sized plume at various angles relative to the subsolar point (0$^\circ${/12h}). These angles include the morning terminator (–90$^\circ${/6h}; minimum sublimation), a morning sector (–30$^\circ${/10h}), the afternoon sector (+30$^\circ${/14h}; maximum sublimation), the evening terminator (+90$^\circ${/18h}), and two nightside sectors (–150$^\circ${/2h}, +150$^\circ${/22h}).
One can clearly see the strong interaction between the sublimated atmosphere and the plume in Figure~\ref{Fig:Plumes}. On the nightside and at the morning terminator (–90$^\circ$), the plume only interacts with the atmosphere in proximity to the plume, with higher temperatures being observable on the plume wings. A canopy shock can also be observed, where the rising plume material interacts with falling particles. While directly around the vent, the sublimated material is still dragged to altitudes of ${\sim}50$~km (winglets), the falling down plume material suppresses the sublimated material to the surface. In the morning (–30$^\circ$) and at the evening terminator (+90$^\circ$), one can observe an asymmetric interaction between material released by volcanic outgassing and sublimation. For both cases, a region of approximately 60$^\circ$ is affected by the plume, which is reflected in higher temperatures of the particles; the atmosphere is enhanced on the side of the plume with more sublimated material. For the case where the plume is positioned in proximity of the point with the highest column density (30$^\circ$), one can see that nearly the full hemisphere (180$^\circ$) is affected by the plume (symmetrically). In summary, the plume influences the sublimated SO$_2$ atmosphere, which in turn influences the flow from a plume, especially on the dayside.

\begin{table*}[h]
\centering
\caption{Dayside and nightside column densities for SO$_2$ in Io’s atmosphere for the interaction of the sublimated SO$_2$ atmosphere with a {medium-sized} plume at different locations. Furthermore, the minimum and maximum column density as well as the values for a purely sublimating atmosphere (subl.) are shown.}
\begin{adjustbox}{max width=\textwidth}
\begin{tabular}{cccccc}
\hline
\hline
Plume    & \multicolumn{4}{c}{Column density [cm$^{-2}$]} \\
Location & $\overline{N}_{\text{dayside}}$ & $\overline{N}_{\text{nightside}}$ & $N_{\text{min}}$ & $N_{\text{max}}$ \\
\hline
–150$^\circ$ & 3.6$\cdot$10$^{16}$ & 15.8$\cdot$10$^{15}$  & 4.9$\cdot$10$^{14}$ & 6.4$\cdot$10$^{17}$ \\ 
–90$^\circ$  & 4.3$\cdot$10$^{16}$ &  8.8$\cdot$10$^{15}$  & 4.8$\cdot$10$^{14}$ & 7.5$\cdot$10$^{17}$ \\ 
–30$^\circ$  & 4.9$\cdot$10$^{16}$ &  2.4$\cdot$10$^{15}$  & 4.5$\cdot$10$^{14}$ & 6.6$\cdot$10$^{17}$ \\ 
+30$^\circ$   & 4.9$\cdot$10$^{16}$ &  2.5$\cdot$10$^{15}$  & 4.6$\cdot$10$^{14}$ & 7.0$\cdot$10$^{17}$ \\ 
+90$^\circ$   & 4.3$\cdot$10$^{16}$ &  8.6$\cdot$10$^{15}$  & 4.4$\cdot$10$^{14}$ & 7.2$\cdot$10$^{17}$ \\ 
+150$^\circ$  & 3.6$\cdot$10$^{16}$ & 15.7$\cdot$10$^{15}$  & 4.7$\cdot$10$^{14}$ & 6.3$\cdot$10$^{17}$ \\
\hline
subl. & 3.6$\cdot$10$^{16}$ &  2.4$\cdot$10$^{15}$  & 5.0$\cdot$10$^{14}$ & 0.9$\cdot$10$^{17}$ \\
\hline
\hline
\end{tabular}
\end{adjustbox}
\label{tab:column_densities_plume}
\end{table*}

\begin{figure*} [t]
    \includegraphics[width=0.99\textwidth]{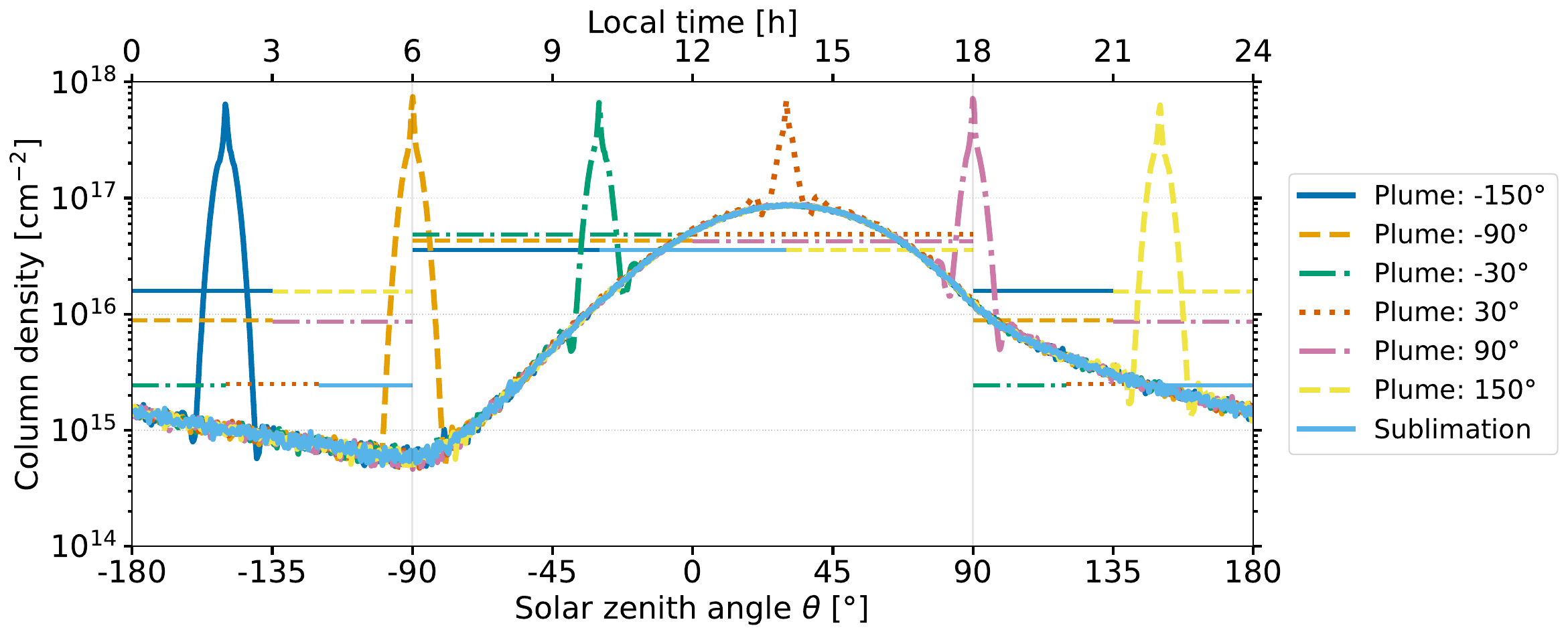}
    \caption{Column density for different positions of the {medium-sized} plume in relation to Io's sublimated SO$_2$ atmosphere, with the same plume locations as in Figure~\ref{Fig:Plumes}. Furthermore, the average dayside and nightside column densities are shown as horizontal lines, also specified in Table~\ref{tab:column_densities_plume}. We find that the {medium-sized} plumes primarily affect the column density in the vicinity of the plume, with winglets of lower column density on both sides of the plume.}
    \label{Fig:Plumes_column_density}
\end{figure*}

The total column density is calculated as the sum of both atmospheric components. In Table~\ref{tab:column_densities_plume}, the dayside and nightside average of the column density, as well as the minimum and maximum values, are shown for the different plume locations and compared to the sublimated atmosphere. The medium-sized plume in our model can significantly increase the column densities, especially on the nightside. In Figure~\ref{Fig:Plumes_column_density}, we show the column densities for the different plume positions compared to the sublimated atmosphere. One can see that the plume mainly affects the column density in the proximity of the plume vent (in contrast to the number densities in Figure~\ref{Fig:Plumes}). Therefore, the interactions between material from the plume and sublimated particles do not seem to affect the total atmospheric mass. However, plumes can significantly alter the local atmosphere composition by displacing sublimated material. Especially when considering different possible plume compositions (e.g., S$_2$, NaCl, KCl), some species in the atmosphere could be enhanced temporarily.
{SO$_2$ column densities from ALMA observations by \citeA{depater2020} show essentially no change in column density between eclipse and sunlight, with typical column densities of $(1.5{\pm}0.3){\cdot}10^{16}$~cm$^{-2}$ and temperatures of $220{-}320$~K, but a large difference in the fractional coverage of SO$_2$. The nearly constant column density and temperature could potentially correspond to a plume on the day- and night-side in Figure~\ref{Fig:Plumes_column_density}.} 

Simulations by \citeA{mcdoniel2017} also find that the local column density is decreased at the interaction of the plume canopies with the sublimated atmosphere. Similarly, we also observe winglets of lower column density on both sides of the plume, which are equivalent to the relatively low-density rings that \citeA{mcdoniel2017} found around the peak column density. These regions can be connected to down-falling plume material from the canopy, interacting with the sublimated atmosphere. However, an asymmetry can also be observed in the size of the winglets, as column density is decreased more on the side where less sublimated material is present. 
\citeA{zhang2003, zhang2004} found that a multiple-ring structure of deposition material could be caused by “bounces” of the falling plume gas during the presence of a sublimated atmosphere, which is related to the sticking coefficient of the gas. In our simulations, we only simulate SO$_2$ with a sticking coefficient of unity and, therefore, do not investigate the multiple-ring structure.

\section{Conclusion}

{In this study, we construct a 2D model of Io's SO$_2$ atmosphere using the DSMC method. To refine our thermal model of the surface temperature,} we conduct a parametric study of thermophysical parameters for the SO$_2$ surface frost by comparing thermal inertia, Bond albedo, and emissivity values obtained from different observation campaigns. Incorporating parameters from the latest observations by \citeA{giles2024} in our thermal model, we find a sublimating SO$_2$ atmosphere with an average column density of $3.5{\cdot}10^{16}$~cm$^{-2}$ on the dayside, which is comparable to observed values.
{Furthermore, we show that the observed asymmetry in SO$_2$ column densities between the sub-Jovian and anti-Jovian hemispheres is likely induced by Io passing through Jupiter's shadow during eclipse, which agrees with results by \citeA{walker2012}.} While more thermal activity has been observed on the anti-Jovian hemisphere, the eclipse reduces the surface temperature on the sub-Jovian hemisphere significantly, reducing the mean dayside column density by a factor of 5.5. {Therefore, spatial variations in SO$_2$  column density between hemispheres could be explained by sublimation alone, while \citeA{tsang2012, tsang2013} and \citeA{giles2024} find that a purely sublimation atmosphere would over-predict the variability with heliocentric distance, calling for a secondary but significant volcanic contribution.}

We also investigate interactions of the sublimated SO$_2$ atmosphere with volcanic plumes, while also accounting for the thermal conductivity of the surface frost. We show that volcanic plumes can significantly alter Io’s local atmosphere by displacing sublimated material, with the effect particularly pronounced when the plume is placed at the point of maximum sublimation, which is shifted 31$^\circ$ from the subsolar point, with particles throughout the hemisphere being impacted. In this study, we focused on modelling SO$_2$ molecules, both for the sublimating and the volcanic component. However,{ besides the main constituent SO$_2$,} the volcanic plumes of Io are expected to consist of a variety of species (e.g., NaCl, KCl, S$_2$). Therefore, the interactions of plume material can change the atmospheric composition, which could also be observable. While our model is designed to capture sublimation and volcanic outgassing of SO$_2$ in Io's atmosphere with reasonable accuracy, it neglects certain factors, such as Io’s three-dimensional structure. Therefore, we are not able to include asymmetric features, such as different vent structures or Io's topography, in our model. Furthermore, the interaction of volcanic plumes of different sizes would be interesting to investigate. However, this lies beyond the scope of our study and is subject to future efforts. 

This study demonstrates that the various processes governing Io's atmosphere are still not fully understood, and fundamental questions remain unanswered. Although ESA's JUpiter ICy moon Explorer \cite<Juice;>{grasset2013} and NASA's Europa Clipper \cite{pappalardo2024} will not come close to Io to allow for in situ measurements, remote sensing observations from Juice and Europa Clipper will provide some valuable data, for example, on Io’s volcanic activity \cite{keane2022}. However, a dedicated mission to the Jovian satellite is necessary to address a wider range of scientific topics and provide more detail, such as the Io Volcano Observer mission concept \cite<IVO;>{hamilton2025}.
With the discovery of so-called lava worlds, a type of terrestrial planet whose planetary surface is mostly or entirely covered by magma oceans or pools of erupted lava \cite{Henning2009}, exploring Io as a laboratory for volcanic processes and the interaction with the atmosphere is relevant not only for Solar System exploration but also for exoplanet research \cite{haslebacher2026}. 
This study emphasises the need for a dedicated in-situ mission to Io to explore this unique world in detail.

\section*{Open Research Section}

The data required to reproduce the figures in the study can be found in \citeA{schlarmann_2025}.

\section*{Conflict of Interest declaration}
The authors declare there are no conflicts of interest for this manuscript.

\acknowledgments
This work has been carried out within the framework of the National Centre of Competence in Research PlanetS supported by the Swiss National Science Foundation under grant 51NF40\_205606. The authors acknowledge the financial support of the SNSF. Part of this work was conducted under the SNSF starting grant 218336.

\appendix

\section{Observations of Io's atmosphere} \label{appendix:observations}

In Table~\ref{tab:observations}, we summarise the column densities of the different  species that have been observed in Io's atmosphere. 

\begin{table*}[h]
\centering
\caption{Observations of different species in the atmosphere of Io.}
\begin{adjustbox}{max width=\textwidth}
\begin{tabular}{ccccccc}
\hline
\hline
Species& Column density [cm$^{-2}$] & Instrument & \multicolumn{2}{c}{Wavelength range}               & Notes        & Reference \\
\hline
SO$_2$ & $4 \cdot 10^{16}$                & IRAM          & mm      & 221.965 GHz         &                    & \citeA{lellouch1990} \\
       & $6 \cdot 10^{17}$                & IRAM/CSO      & mm      & 221.965/143.057 GHz &                    & \citeA{lellouch1992} \\
       & $(0.6{-}1.0)\cdot 10^{16}$/${<}3 \cdot 10^{17}$ & HST/FOS       & UV      & 1980-2300 {\AA}     & trailing           & \citeA{ballester1994} \\
       & ${<}4 \cdot 10^{16}$             & HST/FOS       & UV      & 2270-3300 {\AA}     &                    & \citeA{clarke1994} \\
       & $(5.0{/}7.0)\cdot 10^{15}$/${<}9.3 \cdot 10^{16}$ & HST/GHRS      & UV      & 2095-2135 {\AA}     & leading/trailing   & \citeA{trafton1996} \\
       & $3.7 \cdot 10^{17}$              & HST/WFPC      & near-UV & 2720 {\AA}          &                    & \citeA{spencer1997} \\
       & $(0.70{-}3.25)\cdot 10^{16}$     & HST/FOS       & near-UV & 1900-2300 {\AA}     &                    & \citeA{mcgrath2000} \\
       & $(1{-}4)\cdot 10^{16}$           & HST/STIS      & UV      & 1216 {\AA} (Ly-$\alpha$)  &                    & \citeA{feldman2000} \\
       & $(7{\pm}3)\cdot 10^{16}$         & HST/STIS/WFPC & UV      & 168-306 nm          & Pele               & \citeA{spencer2000} \\
       & $(1{-}2)\cdot 10^{16}$           & HST/STIS      & UV      & 1216 {\AA} (Ly-$\alpha$)  &                    & \citeA{strobel2001} \\
       & $(0.5{-}1.0)\cdot 10^{16}$       & HST/STIS      & UV      & 1479 {\AA}          &                    & \citeA{feaga2002} \\
       & $(1.4{-}2.4)\cdot 10^{17}$       & HST/STIS      & UV      & 2000-3170 {\AA}     & Prometheus         & \citeA{jessup2004} \\
       & $(0.15{-}1.50)\cdot 10^{17}$     & IRTF/TEXES    & mid-IR  & 18.90-18.83 $\mu$m  &                    & \citeA{spencer2005} \\
       & $(1{-}5)\cdot 10^{16}$           & HST/STIS      & UV      & 2000-3170 {\AA}     & Pele               & \citeA{jessup2007} \\
       & ${<}1.5 \cdot 10^{17}$           & IRAM/PdBI     & mm      & 216.643 GHz         &                    & \citeA{moullet2008} \\
       & $(0.018{-}5)\cdot 10^{16}$       & HST/STIS      & UV      & 1216 {\AA} (Ly-$\alpha$)  &                    & \citeA{feaga2009} \\
       & $(2.3{-}4.6)\cdot 10^{16}$       & SMA           & mm      & 346.652/346.523 GHz & leading            & \citeA{moullet2010} \\
       & $(0.7{-}1.1)\cdot 10^{16}$       & SMA           & mm      & 346.652/346.523 GHz & trailing           & \citeA{moullet2010} \\
       & $(0.61{/}1.51)\cdot 10^{17}$     & IRTF/TEXES    & mid-IR  & 19 $\mu$m           & aphel./peri. (anti-Jov.) & \citeA{tsang2012} \\
       & $(0.55{-}1.80)\cdot 10^{17}$     & IRTF/TEXES    & mid-IR  & 19 $\mu$m           &                    & \citeA{tsang2013} \\
       & $(5.5{\pm}0.7)\cdot 10^{15}$     & APEX          & sub mm  & 346.652/346.523 GHz &                    & \citeA{moullet2013} \\
       & $(0.3{-}1.5)\cdot 10^{17}$       & VLT/CRIRES    & near-IR & 4 $\mu$m            &                    & \citeA{lellouch2015} \\
       & $(0.3{-}2.2)\cdot 10^{17}$       & HST/STIS      & near-UV & 2100-2300 {\AA}     &                    & \citeA{jessup2015} \\
       & $(1.5{\pm}0.3)\cdot 10^{16}$     & ALMA          & sub mm  & 333.043-346.652 GHz &                    & \citeA{depater2020} \\
       & $(0.75{-}1.19)\cdot 10^{16}$     & IRAM/NOEMA    & mm      & 258.389-259.599 GHz &                    & \citeA{roth2020} \\
       & ${<}2 \cdot 10^{17}$             & HST/STIS      & UV      & 1216 {\AA} (Ly-$\alpha$)  &                    & \citeA{giono2021} \\
       & $(1.030{\pm}0.032)\cdot 10^{16}$ & ALMA          & sub mm  & 416-432 GHz         & leading            & \citeA{dekleer2024} \\
       & $(3.53{\pm}0.21)\cdot 10^{15}$   & ALMA          & sub mm  & 416-432 GHz         & trailing           & \citeA{dekleer2024} \\
       & $(0.9{/}2.1)\cdot 10^{17}$       & IRTF/TEXES    & mid-IR  & 18.88 $\mu$m        & aphel./peri. (anti-Jov.) & \citeA{giles2024} \\
       & $(1{/}4)\cdot 10^{16}$           & IRTF/TEXES    & mid-IR  & 18.88 $\mu$m        & aphel./peri.  (sub-Jov.) & \citeA{giles2024} \\
SO     & $(2.0{-}6.0)\cdot 10^{14}$ & IRAM          & mm      & 219.949/138.178 GHz &                    & \citeA{lellouch1996} \\
       & $(0.5{-}2.5)\cdot 10^{15}$       & HST/FOS       & near-UV & 1590-2312 {\AA}     &                    & \citeA{mcgrath2000} \\
       & $(1.0{-}7.0)\cdot 10^{14}$       & SMA           & mm      & 346.528 GHz         &                    & \citeA{moullet2010} \\
       & $(3.8{-}4.5)\cdot 10^{14}$       & APEX          & sub mm  & 344.310/346.528 GHz &                    & \citeA{moullet2013} \\
       & $1 \cdot 10^{15}$                & ALMA          & sub mm  & 346.528/344.311 GHz &                    & \citeA{depater2020} \\
S$_2$  & $(1.0{\pm}0.2)\cdot 10^{16}$     & HST/STIS/WFPC & UV      & 168-306 nm          & Pele               & \citeA{spencer2000} \\
       & ${<}7.5 \cdot 10^{14}$           & HST/STIS      & UV      & 2000-3170 {\AA}     & Prometheus         & \citeA{jessup2004} \\
       & $(1.0{-}4.0)\cdot 10^{15}$       & HST/STIS      & UV      & 2000-3170 {\AA}     & Pele               & \citeA{jessup2007} \\
S      & $(0.5{-}1.0)\cdot 10^{14}$       & HST/FOS       & near-UV & 1900 {\AA}          &                    & \citeA{mcgrath2000} \\
       & $(0.36{-}1.70)\cdot 10^{13}$     & HST/STIS      & UV      & 1479 {\AA}          &                    & \citeA{feaga2002} \\
NaCl   & $(0.8{-}20)\cdot 10^{13}$        & IRAM          & mm      & 143.237/234.252 GHz &                    & \citeA{lellouch2003} \\
       & $(1.2{-}1.4)\cdot 10^{13}$       & IRAM/NOEMA    & mm      & 260.233 GHz         &                    & \citeA{roth2020} \\
       & $(0.01{-}1.00)\cdot 10^{14}$     & ALMA          & sub mm  & 338.021/260.223 GHz &                    & \citeA{redwing2022} \\
       & $(5.1{\pm}2.0)\cdot 10^{13}$     & ALMA          & sub mm  & 416-432 GHz         & leading            & \citeA{dekleer2024} \\
       & $(3.3{\pm}1.8)\cdot 10^{13}$     & ALMA          & sub mm  & 416-432 GHz         & trailing           & \citeA{dekleer2024} \\
KCl    & $(3.0{\pm}1.0)\cdot 10^{12}$     & APEX          & sub mm  & 344.820 GHz         &                    & \citeA{moullet2013} \\
       & $(0.01{-}1.00)\cdot 10^{13}$     & ALMA          & sub mm  & 344.820/337.208/260.916 GHz &                    & \citeA{redwing2022} \\
       & $(9.9{\pm}3.9)\cdot 10^{12}$     & ALMA          & sub mm  & 416-432 GHz         & leading            & \citeA{dekleer2024} \\
       & $(3.5{\pm}2.0)\cdot 10^{12}$     & ALMA          & sub mm  & 416-432 GHz         & trailing           & \citeA{dekleer2024} \\
\hline
\hline
\end{tabular}
\end{adjustbox}
\label{tab:observations}
\end{table*}

{\citeA{lellouch1990} conducted the first microwave observation of SO$_2$ from the ground with the 30~m radio telescope of IRAM (Institut de radioastronomie millimétrique), and showed that Io has a global SO$_2$ atmosphere, presenting an averaged column density of $4{\cdot}10^{16}$~cm$^{-2}$.}
\citeA{lellouch1992} performed millimetre-wave observations with IRAM and the 10.4-m CSO (Caltech Submillimeter Observatory) at Mauna Kea, finding a typical partial pressure of $10^{-3}$ Pa and an average column density of $6 \cdot 10^{17}$~cm$^{-2}$. Furthermore, \citeA{lellouch1996} detected SO in Io's atmosphere with IRAM, reporting SO column densities in the range $(2{-}6) \cdot 10^{14}$~cm$^{-2}$ and SO/SO$_2$ mixing ratios varying from $3\%{-}10\%.$ \citeA{lellouch2003} observed NaCl in Io's atmosphere, and determined column densities of $(0.8{-}20)\cdot10^{13}$~cm$^{-2}$, about 0.3\% of SO$_2$. \citeA{moullet2008} performed observations with the IRAM Plateau de Bure Interferometer (PdBI), finding that the equatorial SO$_2$ column density varies strongly with longitude, with a maximum of  $1.5\cdot10^{17}$~cm$^{-2}$. 
\citeA{roth2020} used the NOrthern Extended Millimetre Array (NOEMA) of IRAM to investigate Io's SO$_2$ and NaCl atmosphere, finding column densities of $1.1\cdot10^{16}$ cm$^{-2}$ for SO$_2$ and $(1.2{-}1.4)\cdot10^{13}$ cm$^{-2}$ for NaCl. 
\citeA{moullet2010} performed observations with the Submillimeter Array (SMA), finding averaged SO$_2$ column densities of $(2.3{-}4.6)\cdot 10^{16}$~cm$^{-2}$ on the leading and $(0.7{-}1.1)\cdot10^{16}$~cm$^{-2}$ on the trailing hemisphere. Furthermore, they found SO column densities in the range $(1{-}7)\cdot10^{14}$~cm$^{-2}$, and had low 3-$\sigma$ detections of NaCl. \citeA{moullet2013} used the Atacama Pathfinder EXperiment (APEX) telescope, presenting a tentative detection of KCl and $^{34}$SO$_2$, as well as column densities of SO$_2$ and SO.
\citeA{depater2020} used the Atacama Large Millimeter/submillimeter Array (ALMA) to observe SO$_2$, SO, and KCl when Io went into eclipse, presenting column densities of $(1.5\pm0.3) \cdot 10^{16}$~cm$^{-2}$ for SO$_2$  and $10^{15}$~cm$^{-2}$ for SO. {\citeA{redwing2022} give column densities of $10^{12}{-}10^{14}$~cm$^{-2}$ for NaCl and $10^{11}{-}10^{13}$~cm$^{-2}$ for KCl from ALMA observations.}
Measurements of sulphur isotopes in gaseous SO$_2$, and chlorine isotopes in gaseous NaCl and KCl were performed with ALMA by \citeA{dekleer2024}, presenting isotope ratios $^{34}$S/$^{32}$S${=} (0.0595 \pm 0.0038)$ and $^{37}$Cl/$^{35}$Cl${=}(0.403\pm0.028)$, finding evidence for long-lived volcanism on Io.

\citeA{ballester1994} observed SO$_2$ absorption bands in the ultraviolet (UV) using the Faint Object Spectrograph (FOS) of the Hubble Space Telescope (HST), finding an average column density of $(6{-}10) \cdot 10^{15}$ cm$^{-2}$ for a hemispheric model, and  $3 \cdot 10^{17}$ cm$^{-2}$ for a spatially confined atmosphere ($\sim$~8\% hemispheric areal coverage). \citeA{clarke1994} used HST/FOS to observe SO$_2$ frost reflectivity features during eclipse entry, finding an upper limit of $4 \cdot 10^{16}$ cm$^{-2}$ for the SO$_2$ column density. 
\citeA{mcgrath2000} reported SO$_2$ column densities of $(0.7{-}3.25)\cdot10^{16}$ cm$^{-2}$ from observations made with the HST/FOS of the Pele volcano, the Ra volcano, and a control region, while also detecting SO with column densities of $(0.5{-}2.5)\cdot10^{15}$ cm$^{-2}$ and estimating a S column density of $(5{-}10)\cdot10^{13}$ cm$^{-2}$. 
\citeA{trafton1996} reported UV observations of a SO$_2$ with the Goddard High Resolution Spectrograph (GHRS) on HST, finding a disk-averaged column density of $0.7\cdot 10^{16}$ cm$^{-2}$ and $0.5\cdot 10^{16}$ cm$^{-2}$ for the trailing and leading sunlit hemisphere, respectively, with a maximum SO$_2$ column abundance of $9.3 \cdot 10^{16}$ cm$^{-2}$. \citeA{spencer1997} inferred a SO$_2$ column density of $3.7 \cdot 10^{17}$ cm$^{2}$ from observing the Pele plume with the Wide-Field Planetary Camera 2 (WFPC) of the HST. \citeA{feldman2000} used Lyman-$\alpha$ data obtained with the Space Telescope Imaging Spectrograph (STIS) of HST to map the SO$_2$ distribution, finding SO$_2$ column density of $(1{-}4)\cdot10^{16}$~cm$^{-2}$. \citeA{spencer2000} detected S$_2$ in Io's Pele Plume with HST/WFPC and HST/STIS, with column densities of $(1.0 \pm 0.2)\cdot 10^{16}$~cm$^{-2}$, while reporting SO$_2$ column densities of $(7\pm3) \cdot 10^{16}$~cm$^{-2}$. \citeA{strobel2001} inferred SO$_2$ column densities in the range of $(1-2)\cdot10^{16}$cm$^{-2}$ from the HST/STIS Lyman-$\alpha$ intensities. 
\citeA{feaga2002} constrain the atomic sulphur column density to $(0.36-1.7) \cdot 10^{13}$cm$^{-2}$ using HST/STIS, consistent with a SO$_2$ column density of $(5-10)\cdot10^{15}$~cm$^{-2}$.
\citeA{jessup2004} observed Io’s Prometheus plume with HST/STIS and found SO$_2$ column densities of $1.25 \cdot 10^{17}$~cm$^{-2}$ near the equator, with an additional enhancement of $5 \cdot 10^{16}$~cm$^{-2}$ over Prometheus. They also provide upper limits for S$_2$ based on its non-detection. \citeA{jessup2007} also presented constraints on the composition of the Pele plume, presenting column densities of $(1{-}5)\cdot 10^{16}$~cm$^{-2}$ for SO$_2$ and $(1{-}4)\cdot 10^{15}$~cm$^{-2}$ S$_2$. \citeA{feaga2009} used Lyman-$\alpha$ images obtained with HST/STIS from 1997 to 2001, to derive a global SO$_2$ distribution map, finding a maximum equatorial column density of $5 \cdot 10^{16}$~cm$^{-2}$ at 140$^\circ$ longitude on the anti-Jovian hemisphere. \citeA{roth2014} inferred mixing ratios of atomic oxygen (O) and atomic sulphur (S) relative to SO$_2$ are approximately 10\% and 2\%, respectively.
{\citeA{jessup2015}{, who investigated spatial distribution and diurnal variability of Io's SO$_2$ atmosphere,} 
present SO$_2$ column densities of $(0.3{-}2.2){\cdot}10^{17}$~cm$^{-2}$ from HST/STIS observations.}
\citeA{giono2021} found that HST/STIS images of the surface-reflected Ly-$\alpha$ flux are only sensitive to SO$_2$ column densities between ${\sim}10^{15}$ cm$^{-2}$ and $5\cdot10^{16}$ cm$^{-2}$ due to a strong non-linearity in the relationship between the abundance of SO$_2$ and Ly-$\alpha$ flux. Therefore, they state that previously inferred values via Ly-$\alpha$ observations \cite<e.g.,>{feldman2000, feaga2009} are likely underestimated, inferring a new upper limit of ${\sim}2\cdot10^{17}$ for the SO$_2$ column density.

\citeA{spencer2005} observed Io in the infrared (IR) with TEXES (Texas Echelon Cross Echelle Spectrograph) at the NASA Infrared Telescope Facility (IRTF) 3~m telescopes to infer that the atmospheric SO$_2$ column density varies from about $(0.15{-}1.5)\cdot10^{17}$ cm$^{-2}$. This observation marks the beginning of investigations into the seasonal variations in SO$_2$ column density in Io's atmosphere with TEXES. \citeA{tsang2012} used NASA IRTF/TEXES observations from 2001 to 2010 (over almost a full Jovian year) to derive sub-solar SO$_2$ column densities on the anti-Jovian hemisphere of $0.61\cdot10^{17}$ cm$^{-2}$ near aphelion (5.45 AU) and $1.51\cdot10^{17}$ cm$^{-2}$ approaching perihelion (4.96 AU), suggesting seasonal changes by a factor of $\sim$3.
{\citeA{tsang2013} included IRTF/TEXES observations from 2012 and 2013, constraining a possible volcanic SO$_2$ component to $(6.5{\pm1.0}){\cdot}10^{16}$~cm$^{-2}$}
\citeA{giles2024} extended the previous IRTF/TEXES observations by \citeA{spencer2005}, and \citeA{tsang2012, tsang2013} by data obtained between 2014 and 2023, covering almost two Jovian years. They reporting equatorial maximum column densities between ${\sim}9\cdot10^{16}$ cm$^{-2}$ (aphelion) and ${\sim}2.1\cdot10^{17}$ cm$^{-2}$ (perihelion) on the anti-Jovian hemisphere and between ${\sim}1\cdot10^{16}$ cm$^{-2}$ (aphelion) and ${\sim}4\cdot10^{16}$ cm$^{-2}$ (perihelion) on the sub-Jovian hemisphere (sparse data).
\citeA{lellouch2015} used the cryogenic high-resolution infrared echelle spectrograph (CRIRES) installed on the ESO Very Large Telescope (VLT) and reported SO$_2$ equatorial column densities of $(0.3{-}1.5) \cdot 10^{17}$ cm$^{-2}$. 

\section{Sublimation vapour pressure of SO$_2$}
\label{appendix:vapour_pressure}

\begin{figure}
    \includegraphics[width=0.99\textwidth]{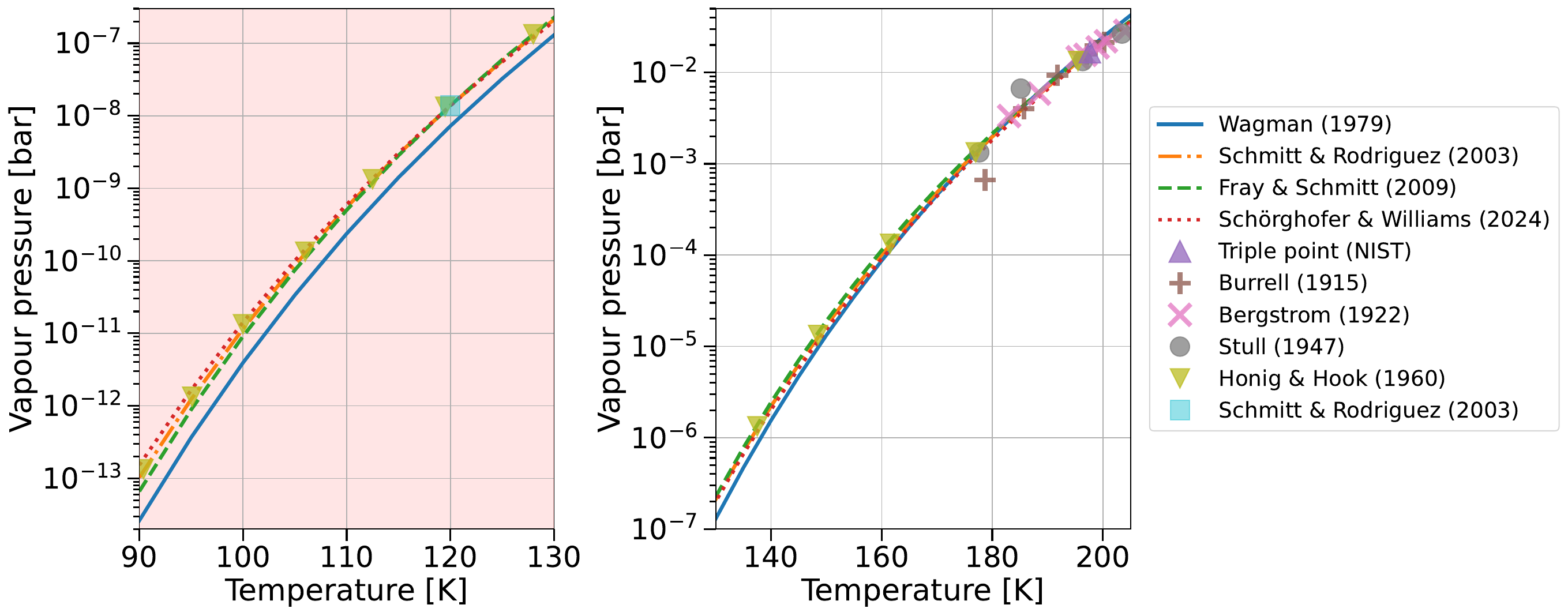}
    \caption{Vapour pressure of solid SO$_2$. Measurement data are from \citeA{burrell1915, bergstrom1922, stull1947, schmitt2003}. Furthermore, thermodynamical curve fits for the vapour pressure from \citeA{wagman1979, schmitt2003,fray2009, schorghofer2024}. Tabulated data are from \citeA{honig1960}, and the NIST triple point temperature ($T_t{=}197.64$ K) and pressure ($p_t{=}1670$ Pa) from \citeA{giauque1938} are shown. The light red area {in the left subplot} represents the range of temperatures measured on Io's surface with \textit{Galileo} PPR \cite{rathbun2004}.}
    \label{Fig:pT}
\end{figure}

Vapour pressure measurements for solid sulphur dioxide (SO$_2$) are rare. In the literature, various approaches have been employed to fit vapour pressure curves to laboratory data.
Based on the Clausius–Clapeyron equation \citeA{wagman1979} gives the following expression (Eq.~\ref{eq:vp_wagman1979}) for the vapour pressure $P_v$ [bar] of SO$_2$, with the vapour pressure coefficients $A{=}1.516{\cdot}10^{8}$~bar and  $B{=}4510$~K. 
\begin{equation}
    P_v = A \cdot e^{-B/T}  \qquad[\text{bar}] \label{eq:vp_wagman1979}
\end{equation}
\citeA{fray2009} fit a polynomial expression (Eq.~\ref{eq:vp_fray2009}) to sublimation pressure data from \citeA{burrell1915}, \citeA{bergstrom1922}, and \citeA{schmitt2003}, to calculate the vapour pressure $P_v$ [bar] of SO$_2$ in the temperature range between {$15$~K and $197.63$~K}, with $A_0{=}15.6, A_1{=}{-}3508, A_2{=}{-}9.401{\cdot}10^{4}, A_3{=}4.152{\cdot}10^{6}, A_4{=}{-}6.946{\cdot}10^{7}$. 
\begin{equation}
    \ln \left(P_v\right) = A_0 + \sum_{i=1}^n \frac{A_i}{T^i}  \label{eq:vp_fray2009} \qquad[\text{bar}]
\end{equation}
The data point from \citeA{schmitt2003} ($P_v{=}(9.0{\pm}0.25){\cdot}10^{-9}$~bar at $120$~K)  originates from a thermodynamical curve fit of the tabulated data of \citeA{honig1960} 
for their sublimation rate measurement. \citeA{honig1960} employed {a so-called \textit{computing machine}, presumably one of the first basic computers,} to determine the temperatures corresponding to specific pressures from a set of experimental values, which mainly originate from \citeA{stull1947}. \citeA{schmitt2003} derived a thermodynamical curve fit (Eq.~\ref{eq:vp_schmitt2003}) of the tabulated data from \citeA{honig1960} for the saturation vapour pressure $P_v$ [bar] of SO$_2$.
\begin{equation}
    P_v = 2.0 \cdot 10^6 \cdot \sqrt{T} \cdot e^{4200/T} \qquad[\text{bar}] \label{eq:vp_schmitt2003}
\end{equation}
\citeA{schorghofer2024} presented a fit for the temperature-dependent vapour pressure $P_v$ [Pa] of SO$_2$ based on data from \citeA{giauque1938}, \citeA{stull1947}, and \citeA{schmitt2003}. They perform a least-squares fit to the logarithm of the vapour pressures to obtain the coefficients $b_0{=}9{\pm}14, b_1{=}3775{\pm}361, b_2{=}3{\pm}2$ for the Rankine-Dupre formula: 
\begin{equation}
    \ln(P_v) = b_0 - \frac{b_1}{T} + b_2 \ln T \qquad[\text{Pa}]
\end{equation}
However, the coefficients in \citeA{schorghofer2024} are rounded too much, leading to strong deviations from the measurements and the other curves when using the coefficients above. \citeA{schorghofer2024} privately communicated the unrounded coefficients $b_0{=}9.435, b_1{=}3775.1, b_2{=}3.2248$ from their curve fit.

In Figure \ref{Fig:pT} we show an overview of the measurement data and the derived thermodynamic curve fits for the vapour pressure of SO$_2$. Especially at low temperatures, the thermodynamic curve fit from \citeA{wagman1979} differs significantly from the results of the other curve fits (e.g., a 57\% deviation at 100~K and a 47\% deviation at 120~K from \citeA{fray2009}). Furthermore, \citeA{schmitt2003} state that the vapour pressure curve of \citeA{wagman1979} yields a too high value at the triple point (+12\%), which casts some doubt on its accuracy. Therefore, we decided to use the vapour pressure curves of \citeA{fray2009} for our simulations.

\section{Parameter study of thermophysical parameters} \label{sec:Parameter_study}

\begin{figure*}
    \includegraphics[width=0.99\textwidth]{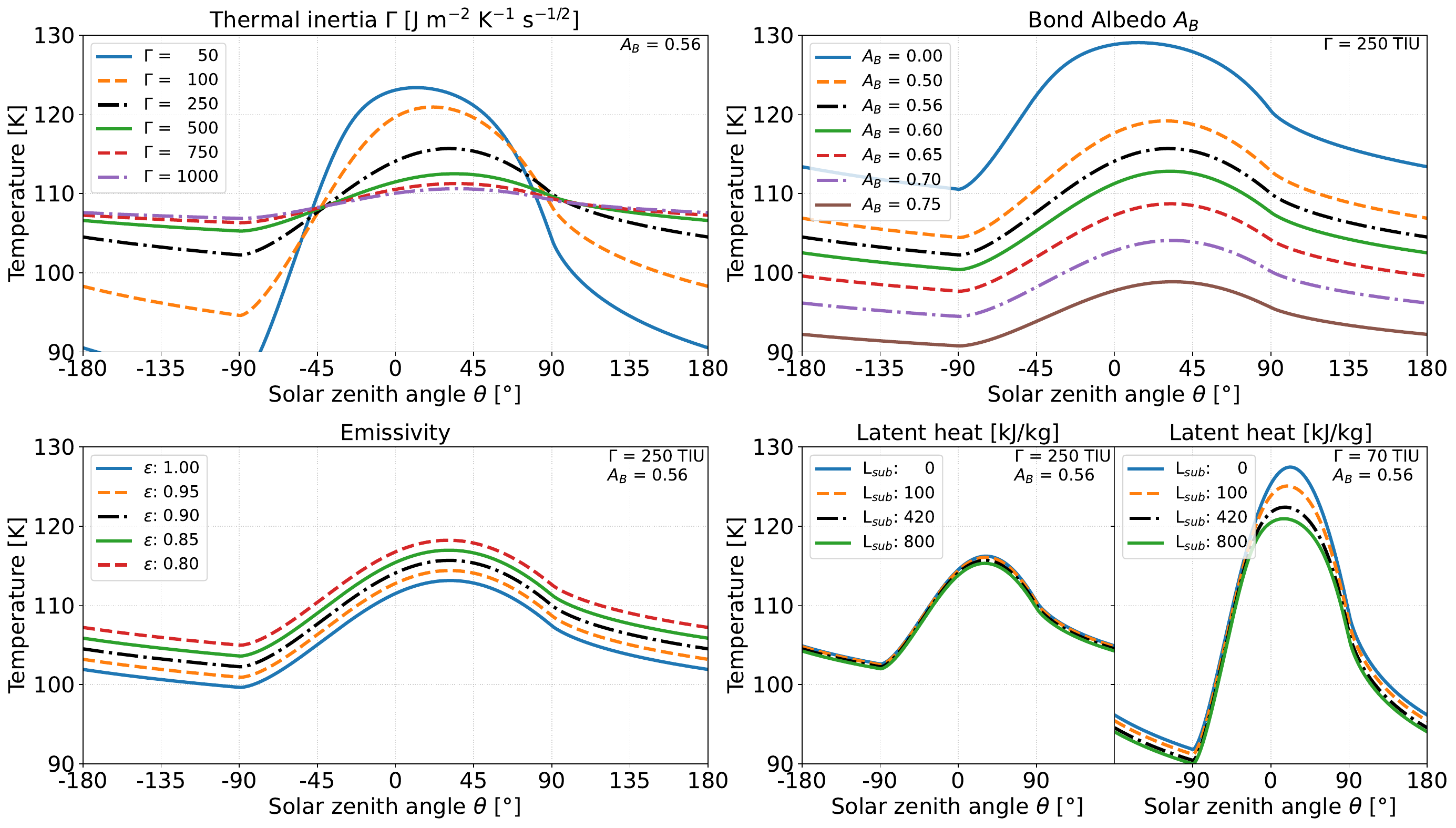}
    \caption{Dependence of the temperature profile on different thermophysical parameters. The reference atmosphere is modeled with thermal inertia $\Gamma{=}250$~\tiunit, a Bond albedo $A_B{=}0.56$, an emissivity $\epsilon{=}0.9$, and a latent heat of sublimation $L_{\text{sub}}{=}420$~kJ/kg as default values, {represented by the black dash–dotted line}.}
    \label{Fig:comparison_thermal_parameters}
\end{figure*}

In Figure \ref{Fig:comparison_thermal_parameters}, we compare the influence of different thermal parameters on the temperature profile. One can see that the thermal inertia $\Gamma$ ($50{-}1000$~\tiunit) and the Bond albedo  $A_B$ (0--0.75) have a strong influence on the temperature profile characteristics. While the emissivity $\epsilon$ (0.8--1.0) does not change the shape of the {surface} temperature profile, a higher emissivity decreases the temperature. The latent heat of sublimation of SO$_2$ (${L_{\text{sub}}\sim}$420~kJ/kg) does not significantly alter the temperature profile for the thermal inertia used in this study. However, for low thermal inertias (e.g., 70~\tiunit), similar values can significantly alter the temperature profile. Furthermore, our reference atmosphere, calculated with the thermophysical parameters of \citeA{giles2024}, is shown as a black dash-dotted line.

\section{Number densities of SO$_2$ and O$_2$} \label{sec:Reference_atm}

\begin{figure*}[h!] 
    \includegraphics[width=0.99\textwidth]{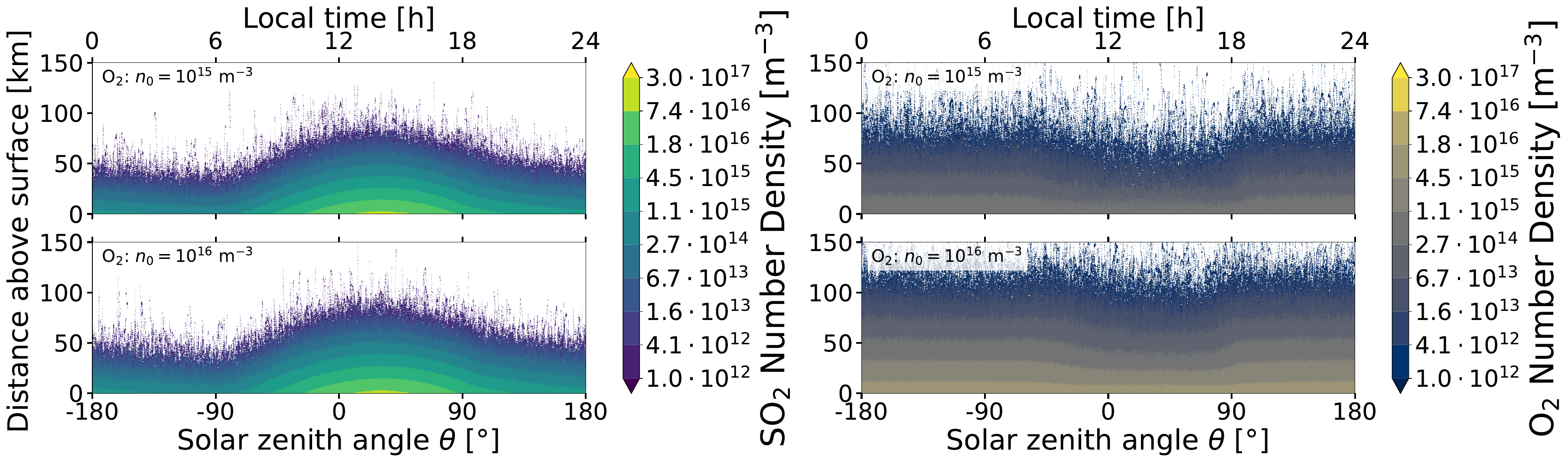}
    \caption{{Influence of a background O$_2$ atmosphere with different surface densities of O$_2$ ($10^{15}$~m$^{-3}$, $10^{16}$~m$^{-3}$) on the reference SO$_2$ atmosphere calculated with the thermophysical parameters from \citeA{giles2024}. In Figure~\ref{Fig:O2}, the corresponding temperatures and zonal velocity are shown.}}
    \label{Fig:reference_atm}
\end{figure*}

In Figure~\ref{Fig:reference_atm}, the number densities of SO$_2$ and O$_2$ are shown against both solar zenith angle and local time, while varying the O$_2$ surface density from $10^{15}$~m$^{-3}$ to $10^{16}$~m$^{-3}$, corresponding to the temperatures and velocities in Figure~\ref{Fig:O2}. The number density of sublimated SO$_2$ in Io's atmosphere, which is calculated using thermophysical parameters from \citeA{giles2024}, is not altered by the introduction of a background O$_2$ atmosphere.

%
%

\bibliography{Literatur}

@article{ackley2021,
  title={Hybrid dust-tracking method for modeling {I}o's {Tvashtar} volcanic plume},
  author={Ackley, Peter C and Hoey, William A and Trafton, Laurence M and Goldstein, David B and Varghese, Philip L},
  journal={Icarus},
  volume={359},
  pages={114274},
  year={2021},
  publisher={Elsevier}
}

@article{austin2000,
  title={Rarefied gas model of {I}o's sublimation-driven atmosphere},
  author={Austin, J Victor and Goldstein, David B},
  journal={Icarus},
  volume={148},
  number={2},
  pages={370--383},
  year={2000},
  publisher={Elsevier}
}

@article{bagenal2020,
author = {Bagenal, Fran and Dols, Vincent},
title = {The Space Environment of {Io} and {Europa}},
journal = {Journal of Geophysical Research: Space Physics},
volume = {125},
number = {5},
pages = {e2019JA027485},
doi = {https://doi.org/10.1029/2019JA027485},
year = {2020}
}

@article{bart2004,
title = {Ridges and tidal stress on Io},
journal = {Icarus},
volume = {169},
number = {1},
pages = {111-126},
year = {2004},
note = {Special Issue: Io after Galileo},
issn = {0019-1035},
doi = {https://doi.org/10.1016/j.icarus.2004.01.003},
url = {https://www.sciencedirect.com/science/article/pii/S0019103504000375},
author = {Gwendolyn D. Bart and Elizabeth P. Turtle and Windy L. Jaeger and Laszlo P. Keszthelyi and Richard Greenberg}
}

@article{bellucci2004,
  title={{Cassini/VIMS} observation of an {Io} post-eclipse brightening event},
  author={Bellucci, G and D'Aversa, E and Formisano, V and Cruikshank, D and Nelson, RM and Clark, RN and Baines, KH and Matson, D and Brown, RH and McCord, TB and others},
  journal={Icarus},
  volume={172},
  number={1},
  pages={141--148},
  year={2004},
  publisher={Elsevier}
}

@article{bergstrom1922,
  title={The vapor pressure of sulfur dioxide and ammonia},
  author={Bergstrom, FW},
  journal={The Journal of Physical Chemistry},
  volume={26},
  number={4},
  pages={358--376},
  year={1922},
  publisher={ACS Publications}
}

@article{binder1964,
	title = {Evidence for an atmosphere on {Io}},
	volume = {3},
	copyright = {https://www.elsevier.com/tdm/userlicense/1.0/},
	issn = {00191035},
	doi = {10.1016/0019-1035(64)90038-7},
	language = {en},
	number = {4},
	journal = {Icarus},
	author = {Binder, A and Cruikshank, D},
	month = nov,
	year = {1964},
	pages = {299--305},
}

@book{bird1994,
	address = {Oxford : New York},
	series = {Oxford engineering science series},
	title = {Molecular gas dynamics and the direct simulation of gas flows},
	isbn = {978-0-19-856195-8},
	language = {en},
	number = {42},
	publisher = {Clarendon Press ; Oxford University Press},
	author = {Bird, G. A.},
	year = {1994}
}

@article{burrell1915,
  title={The Vapor Pressures of Sulfur Dioxide and Nitrous Oxide at Temperatures Below Their Normal Boiling Points.},
  author={Burrell, GA and Robertson, IW},
  journal={Journal of the American Chemical Society},
  volume={37},
  number={12},
  pages={2691--2694},
  year={1915},
  publisher={ACS Publications}
}

@article{brown1981,
  title={The {Jupiter} hot plasma torus-Observed electron temperature and energy flows},
  author={Brown, Robert A},
  journal={Astrophysical Journal, Part 1, vol. 244, Mar. 15, 1981, p. 1072-1080.},
  volume={244},
  pages={1072--1080},
  year={1981}
}

@article{carlson1997,
  title={The distribution of sulfur dioxide and other infrared absorbers on the surface of {Io}},
  author={Carlson, RW and Smythe, WD and Lopes-Gautier, RMC and Davies, AG and Kamp, LW and Mosher, JA and Soderblom, LA and Leader, FE and Mehlman, R and Clark, Roger N and others},
  journal={Geophysical research letters},
  volume={24},
  number={20},
  pages={2479--2482},
  year={1997},
  publisher={Wiley Online Library}
}

@article{cruikshank1973,
  title={The post-eclipse brightening of {Io}},
  author={Cruikshank, Dale P and Murphy, Robert E},
  journal={Icarus},
  volume={20},
  number={1},
  pages={7--17},
  year={1973},
  publisher={Elsevier}
}

@article{cruikshank2010,
  title={Eclipse reappearances of {Io}: {T}ime-resolved spectroscopy (1.9--4.2 $\mu$m)},
  author={Cruikshank, Dale P and Emery, Joshua P and Kornei, Katherine A and Bellucci, Giancarlo and d’Aversa, Emiliano},
  journal={Icarus},
  volume={205},
  number={2},
  pages={516--527},
  year={2010},
  publisher={Elsevier}
}

@article{doute2001,
  title={Mapping {SO$_2$} frost on {Io} by the modeling of {NIMS} hyperspectral images},
  author={Dout{\'e}, Sylvain and Schmitt, Bernard and Lopes-Gautier, Rosaly and Carlson, Robert and Soderblom, Laurence and Shirley, James and {Galileo} {NIMS} Team and others},
  journal={Icarus},
  volume={149},
  number={1},
  pages={107--132},
  year={2001},
  publisher={Elsevier}
}

@article{dott2025,
  title={Observed Latitudinal, Longitudinal and Temporal Variability of {Io}'s Atmosphere Simulated by a Purely Sublimation Driven Atmosphere},
  author={Dott, A-C and Saur, J and Schlegel, S and Strobel, DF and de Kleer, K and de Pater, I},
  journal={Journal of Geophysical Research: Planets},
  volume={130},
  number={7},
  pages={e2024JE008869},
  year={2025},
  publisher={Wiley Online Library}
}

@article{dundas2017,
  title={Effects of lava heating on volatile-rich slopes on {Io}},
  author={Dundas, Colin M},
  journal={Journal of Geophysical Research: Planets},
  volume={122},
  number={3},
  pages={546--559},
  year={2017},
  publisher={Wiley Online Library}
}

@ARTICLE{ferrari2018,
       author = {{Ferrari}, C.},
        title = "{Thermal Properties of Icy Surfaces in the Outer Solar System}",
      journal = {Space Science Reviews},
         year = 2018,
        month = dec,
       volume = {214},
       number = {8},
          eid = {111},
        pages = {111},
          doi = {10.1007/s11214-018-0546-x},
     publisher={Springer}
}

@article{fray2009,
  title={Sublimation of ices of astrophysical interest: A bibliographic review},
  author={Fray, N and Schmitt, Bernard},
  journal={Planetary and Space Science},
  volume={57},
  number={14-15},
  pages={2053--2080},
  year={2009},
  publisher={Elsevier}
}

@article{geissler2004,
  title={Cassini observations of {Io}'s visible aurorae},
  author={Geissler, Paul and McEwen, Alfred and Porco, Carolyn and Strobel, Darrell and Saur, Joachim and Ajello, Joseph and West, Robert},
  journal={Icarus},
  volume={172},
  number={1},
  pages={127--140},
  year={2004},
  publisher={Elsevier}
}

@article{giono2021,
  title={{Io}'s {SO$_{2}$} atmosphere from {HST} {Lyman}-$\alpha$ images: 1997 to 2018},
  author={Giono, Gabriel and Roth, Lorenz},
  journal={Icarus},
  volume={359},
  pages={114212},
  year={2021},
  publisher={Elsevier}
}

@article{giauque1938,
  title={Sulfur dioxide. The heat capacity of solid and liquid. Vapor pressure. Heat of vaporization. The entropy values from thermal and molecular data},
  author={Giauque, WF and Stephenson, CC},
  journal={Journal of the American Chemical Society},
  volume={60},
  number={6},
  pages={1389--1394},
  year={1938},
  publisher={ACS Publications}
}

@article{grasset2013,
  title={{JUpiter} {ICy} moons {Explorer} ({JUICE}): An {ESA} mission to orbit {G}anymede and to characterise the {Jupiter} system},
  author={Grasset, Olivier and Dougherty, MK and Coustenis, A and Bunce, EJ and Erd, C and Titov, D and Blanc, M and Coates, A and Drossart, P and Fletcher, LN and others},
  journal={Planetary and Space Science},
  volume={78},
  pages={1--21},
  year={2013},
  publisher={Elsevier}
}

@book{honig1960,
  title={Vapor pressure data for some common gases},
  author={Honig, Richard E and Hook, Harvey O},
  year={1960},
  publisher={David Sarnoff Research Center, RCA}
}

@article{lellouch2007,
  title={Io’s atmosphere},
  author={Lellouch, Emmanuel and McGrath, Melissa A. and Jessup, Kandis Lea},
  journal={Io After {Galileo}},
  pages={231--264},
  year={2007},
  publisher={Springer}
}

@article{hanel1979,
  title={Infrared observations of the {Jovian} system from {Voyager} 1},
  author={Hanel, R and Conrath, B and Flasar, M and Kunde, V and Lowman, P and Maguire, W and Pearl, J and Pirraglia, J and Samuelson, R and Gautier, D and others},
  journal={Science},
  volume={204},
  number={4396},
  pages={972--976},
  year={1979},
  publisher={American Association for the Advancement of Science}
}

@incollection{haslebacher2026,
  author       = {C. Haslebacher and E. Bolmont and M. Cilibrasi and J. Grone and N. Haslebacher and R. Helled and M. Kervazo and N. Ligterink and C. Lovis and L. Mayer and R. Ottersberg and A. Oza and C. Patty and A. Pommerol and G. Portyankina and A. Rhoden and L. Schlarmann and Y. Shibaike and V. Singh and A. Vorburger and P. Wurz},
  title={Active moons in our {S}olar {S}ystem and beyond--{I}o, {E}uropa, {E}nceladus, {T}riton, and exomoons},
  booktitle    = {The {N}ational {C}enter for {C}ompetence in {R}esearch, {P}lanet{S}: A {S}wiss-wide network expanding planetary sciences},
  publisher    = {Springer},
  year         = {2026},
  editor       = {Benz, W. and others},
  note         = {Accepted for publication in the NCCR PlanetS Legacy Book. Preprint available at \url{https://arxiv.org/abs/2604.12104}}
}

@article{Henning2009,
   doi = {10.1088/0004-637X/707/2/1000},
   year = {2009},
   month = {dec},
   publisher = {The American Astronomical Society},
   volume = {707},
   number = {2},
   pages = {1000},
   author = {Henning, Wade G. and O'Connell, Richard J. and Sasselov, Dimitar D.},
   title = {Tidally Heated Terrestrial Exoplanets: {Viscoelastic} Response Models},
   journal = {The Astrophysical Journal}
}

@article{ingersoll1985,
  title={Supersonic meteorology of {Io}: Sublimation-driven flow of {SO$_2$}},
  author={Ingersoll, Andrew P and Summers, Michael E and Schlipf, Steve G},
  journal={Icarus},
  volume={64},
  number={3},
  pages={375--390},
  year={1985},
  publisher={Elsevier}
}

@inproceedings{ingersoll1993,
  title={A Nightside Atmosphere of {Io}?},
  author={Ingersoll, Andrew P and Lebeau Jr., RP},
  booktitle = {{AAS/D}ivision for {P}lanetary {S}ciences Meeting Abstracts \#25},
  series = {{AAS/D}ivision for {P}lanetary {S}ciences Meeting Abstracts},
  volume={25},
  pages={1075},
  month = jun,
  year={1993}
}

@article{jessup2015,
  title={Spatially resolved {HST/STIS} observations of {Io}’s dayside equatorial atmosphere},
  author={Jessup, Kandis Lea and Spencer, John R},
  journal={Icarus},
  volume={248},
  pages={165--189},
  year={2015},
  publisher={Elsevier}
}

@article{johnson1995,
	title = {Stealth plumes on {Io}},
	volume = {22},
	issn = {0094-8276, 1944-8007},
	url = {https://agupubs.onlinelibrary.wiley.com/doi/10.1029/95GL03084},
	doi = {10.1029/95GL03084},
	language = {en},
	number = {23},
	urldate = {2025-05-31},
	journal = {Geophysical Research Letters},
	author = {Johnson, Torrence V. and Matson, Dennis L. and Blaney, Diana L. and Veeder, Glenn J. and Davies, Ashley},
	month = dec,
	year = {1995},
	pages = {3293--3296},
}

@article{kerton1996,
  title={The state of {SO$_2$} on {Io}'s surface},
  author={Kerton, C.R. and Fanale, F.P. and Salvail, J.R.},
  journal={Journal of Geophysical Research: Planets},
  volume={101},
  number={E3},
  pages={7555--7563},
  year={1996},
  publisher={Wiley Online Library}
}

@article{koura1991,
  title={Variable soft sphere molecular model for inverse-power-law or {Lennard-Jones} potential},
  author={Koura, Katsuhisa and Matsumoto, Hiroaki},
  journal={Physics of fluids A: fluid dynamics},
  volume={3},
  number={10},
  pages={2459--2465},
  year={1991},
  publisher={American Institute of Physics}
}

@article{laver2008,
  title={Spatially resolved {SO$_2$} ice on {Io}, observed in the near {IR}},
  author={Laver, Conor and de Pater, Imke},
  journal={Icarus},
  volume={195},
  number={2},
  pages={752--757},
  year={2008},
  publisher={Elsevier}
}

@article{laver2009,
  title={The global distribution of sulfur dioxide ice on {Io}, observed with {OSIRIS} on the {WM} {Keck} telescope},
  author={Laver, Conor and de Pater, Imke},
  journal={Icarus},
  volume={201},
  number={1},
  pages={172--181},
  year={2009},
  publisher={Elsevier}
}

@article{matson1981,
  title={Heat flow from {Io} ({JI})},
  author={Matson, Dennis L and Ransford, Gary A and Johnson, Torrence V},
  journal={Journal of Geophysical Research: Solid Earth},
  volume={86},
  number={B3},
  pages={1664--1672},
  year={1981},
  publisher={Wiley Online Library}
}

@article{mcdonald2022,
  title={Aeolian sediment transport on {Io} from lava--frost interactions},
  author={McDonald, George D and M{\'e}ndez Harper, Joshua and Ojha, Lujendra and Corlies, Paul and Dufek, Josef and Ewing, Ryan C and Kerber, Laura},
  journal={Nature communications},
  volume={13},
  number={1},
  pages={2076},
  year={2022},
  publisher={Nature Publishing Group UK London}
}

@article{mcdoniel2015,
  title={Three-dimensional simulation of gas and dust in {Io}’s {Pele} plume},
  author={McDoniel, William J and Goldstein, David B and Varghese, Philip L and Trafton, Laurence M},
  journal={Icarus},
  volume={257},
  pages={251--274},
  year={2015},
  publisher={Elsevier}
}

@article{mcdoniel2017,
  title={The interaction of {Io}’s plumes and sublimation atmosphere},
  author={McDoniel, William J and Goldstein, David B and Varghese, Philip L and Trafton, Laurence M},
  journal={Icarus},
  volume={294},
  pages={81--97},
  year={2017},
  publisher={Elsevier}
}

@article{mcdoniel2019,
  title={Simulation of {Io}’s plumes and {Jupiter}’s plasma torus},
  author={McDoniel, William J and Goldstein, David B and Varghese, Philip L and Trafton, Laurence M},
  journal={Physics of Fluids},
  volume={31},
  number={7},
  year={2019},
  publisher={AIP Publishing}
}

@article{mcewen1983,
  title={Two classes of volcanic plumes on {Io}},
  author={McEwen, Alfred S and Soderblom, Laurence A},
  journal={Icarus},
  volume={55},
  number={2},
  pages={191--217},
  year={1983},
  publisher={Elsevier}
}

@article{mcewen1988,
  title={The global distribution, abundance, and stability of {SO$_2$} on {Io}},
  author={McEwen, Alfred S and Johnson, Torrence V and Matson, Dennis L and Soderblom, Laurence A},
  journal={Icarus},
  volume={75},
  number={3},
  pages={450--478},
  year={1988},
  publisher={Elsevier}
}

@article{milazzo2001,
  title={Observations and initial modeling of lava-{SO$_2$} interactions at {Prometheus}, {Io}},
  author={Milazzo, MP and Keszthelyi, LP and McEwen, AS},
  journal={Journal of Geophysical Research: Planets},
  volume={106},
  number={E12},
  pages={33121--33127},
  year={2001},
  publisher={Wiley Online Library}
}

@article{morabito1979,
  title={Discovery of currently active extraterrestrial volcanism},
  author={Morabito, LA and Synnott, SP and Kupferman, PN and Collins, Stewart A},
  journal={Science},
  volume={204},
  number={4396},
  pages={972--972},
  year={1979},
  publisher={American Association for the Advancement of Science}
}

@article{moreno1991,
  title={{Io}'s volcanic and sublimation atmospheres},
  author={Moreno, Miguel A and Schubert, Gerald and Baumgardner, John and Kivelson, Margaret G and Paige, David A},
  journal={Icarus},
  volume={93},
  number={1},
  pages={63--81},
  year={1991},
  publisher={Elsevier}
}

@article{morrison1980,
  title={Io: Observational constraints on internal energy and thermophysics of the surface},
  author={Morrison, David and Telesco, CM},
  journal={Icarus},
  volume={44},
  number={2},
  pages={226--233},
  year={1980},
  publisher={Elsevier}
}

@inproceedings{nash1980,
  title={Periodicity in {Io}'s atmospheric mass: Evidence from post-eclipse brightness},
  author={Nash, DB and Matson, DL},
  booktitle={Lunar and {P}lanetary {S}cience {XI}, P. 794-796. Abstract.},
  volume={11},
  pages={794--796},
  year={1980}
}

@article{pappalardo2024,
  title={Science overview of the {Europa} {C}lipper {M}ission},
  author={Pappalardo, Robert T and Buratti, Bonnie J and Korth, Haje and Senske, David A and Blaney, Diana L and Blankenship, Donald D and Burch, James L and Christensen, Philip R and Kempf, Sascha and Kivelson, Margaret G and others},
  journal={Space Science Reviews},
  volume={220},
  number={4},
  pages={40},
  year={2024},
  publisher={Springer}
}

@article{peale1979,
  title={Melting of {Io} by tidal dissipation},
  author={Peale, Stanton J and Cassen, P and Reynolds, Ray T},
  journal={Science},
  volume={203},
  number={4383},
  pages={892--894},
  year={1979},
  publisher={American Association for the Advancement of Science}
}

@article{persad2016,
  title={Expressions for the evaporation and condensation coefficients in the {H}ertz-{K}nudsen relation},
  author={Persad, Aaron H and Ward, Charles A},
  journal={Chemical reviews},
  volume={116},
  number={14},
  pages={7727--7767},
  year={2016},
  publisher={ACS Publications}
}

@article{roth2014,
  title={A phenomenological model of {Io}’s {UV} aurora based on {HST/STIS} observations},
  author={Roth, Lorenz and Saur, Joachim and Retherford, Kurt D and Feldman, Paul D and Strobel, Darrell F},
  journal={Icarus},
  volume={228},
  pages={386--406},
  year={2014},
  publisher={Elsevier}
}

@article{summers1996,
	title = {Photochemistry and {Vertical} {Transport} in {Io}'s {Atmosphere} and {Ionosphere}},
	volume = {120},
	issn = {00191035},
	url = {https://linkinghub.elsevier.com/retrieve/pii/S0019103596900512},
	doi = {10.1006/icar.1996.0051},
	language = {en},
	number = {2},
	urldate = {2025-06-01},
	journal = {Icarus},
	author = {Summers, Michael E. and Strobel, Darrell F.},
	month = apr,
	year = {1996},
	pages = {290--316},
}

@article{smyth2004,
	title = {Impact of electron chemistry on the structure and composition of {Io}'s atmosphere},
	volume = {171},
	copyright = {https://www.elsevier.com/tdm/userlicense/1.0/},
	issn = {00191035},
	url = {https://linkinghub.elsevier.com/retrieve/pii/S0019103504000995},
	doi = {10.1016/j.icarus.2004.04.001},
	language = {en},
	number = {1},
	urldate = {2025-05-30},
	journal = {Icarus},
	author = {Smyth, William H. and Wong, M.C.},
	month = sep,
	year = {2004},
	pages = {171--182},
}

@article{thelen2024,
  title={Io’s {SO$_2$} and {NaCl} Wind Fields from {ALMA}},
  author={Thelen, Alexander E. and de Kleer, Katherine and Cordiner, Martin A. and de Pater, Imke and Moullet, Arielle and Luszcz-Cook, Statia},
  journal={The Astrophysical Journal Letters},
  volume={978},
  number={1},
  pages={L1},
  year={2024},
  publisher={IOP Publishing}
}

@article{ballester1994,
  title={Detection of the {SO$_2$} atmosphere on {Io} with the {Hubble} {Space} {Telescope}},
  author={Ballester, Gilda E and McGrath, MA and Strobel, DF and Zhu, Xun and Feldman, Paul D and Moos, HW},
  journal={Icarus},
  volume={111},
  number={1},
  pages={2--17},
  year={1994},
  publisher={Elsevier}
}

@article{dekleer2024,
  title={Isotopic evidence of long-lived volcanism on {Io}},
  author={de Kleer, Katherine and Hughes, Ery C and Nimmo, Francis and Eiler, John and Hofmann, Amy E and Luszcz-Cook, Statia and Mandt, Kathy},
  journal={Science},
  volume={384},
  number={6696},
  pages={682--687},
  year={2024},
  publisher={American Association for the Advancement of Science}
}

@article{depater2020,
  title={ALMA observations of {Io} going into and coming out of eclipse},
  author={{de Pater}, Imke and Luszcz-Cook, Statia and Rojo, Patricio and Redwing, Erin and De Kleer, Katherine and Moullet, Arielle},
  journal={The Planetary Science Journal},
  volume={1},
  number={3},
  pages={60},
  year={2020},
  publisher={IOP Publishing}
}

@article{depater2020spat,
  title={High spatial and spectral resolution observations of the forbidden 1.707 $\mu$m rovibronic {SO} emissions on {Io}: Evidence for widespread stealth volcanism},
  author={{de Pater}, Imke and de Kleer, Katherine and {\'A}d{\'a}mkovics, M{\'a}t{\'e}},
  journal={The Planetary Science Journal},
  volume={1},
  number={2},
  pages={29},
  year={2020b},
  publisher={IOP Publishing}
}

@ARTICLE{depater2021,
       author = {{de Pater}, Imke and {Keane}, James T. and {de Kleer}, Katherine and {Davies}, Ashley Gerard},
        title = "{A 2020 Observational Perspective of Io}",
      journal = {Annual Review of Earth and Planetary Sciences},
         year = 2021,
        month = may,
       volume = {49},
          doi = {10.1146/annurev-earth-082420-095244},
       adsurl = {https://ui.adsabs.harvard.edu/abs/2021AREPS..49..643D}
}

@article{depater2023,
  title={The plumes and atmosphere of {Io}},
  author={{de Pater}, Imke and Goldstein, David and Lellouch, Emmanuel},
  journal={Io: A New View of {Jupiter}’s Moon},
  pages={233--290},
  year={2023},
  publisher={Springer}
}

@article{clarke1994,
  title={{Hubble} {Space} {Telescope} {UV} spectral observations of {Io} passing into eclipse},
  author={Clarke, John T and Ajello, Joe and Luhmann, Janet and Schneider, Nick and Kanik, Isik},
  journal={Journal of Geophysical Research: Planets},
  volume={99},
  number={E4},
  pages={8387--8402},
  year={1994},
  publisher={Wiley Online Library}
}

@article{feaga2002,
  title={The abundance of atomic sulfur in the atmosphere of {Io}},
  author={Feaga, Lori M and McGrath, Melissa A and Feldman, Paul D},
  journal={The Astrophysical Journal},
  volume={570},
  number={1},
  pages={439},
  year={2002},
  publisher={IOP Publishing}
}

@article{feaga2009,
  title={{Io}'s dayside {SO$_2$} atmosphere},
  author={Feaga, Lori M and McGrath, Melissa and Feldman, Paul D},
  journal={Icarus},
  volume={201},
  number={2},
  pages={570--584},
  year={2009},
  publisher={Elsevier}
}

@article{feldman2000,
  title={{Lyman}-$\alpha$ imaging of the {SO$_2$} distribution on {Io}},
  author={Feldman, Paul D and Strobel, Darrell F and Moos, H Warren and Retherford, Kurt D and Wolven, Brian C and McGrath, Melissa A and Roesler, Fred L and Woodward, R Carey and Oliversen, Ronald J and Ballester, Gilda E},
  journal={Geophysical research letters},
  volume={27},
  number={12},
  pages={1787--1790},
  year={2000},
  publisher={Wiley Online Library}
}

@article{gerig2018,
  title={On deviations from free-radial outflow in the inner coma of comet 67P/Churyumov--Gerasimenko},
  author={Gerig, S-B and Marschall, Raphael and Thomas, Nicholas and Bertini, Ivano and Bodewits, Dennis and Davidsson, Bj{\"o}rn and Fulle, Marco and Ip, W-H and Keller, Horst Uwe and K{\"u}ppers, Michael and others},
  journal={Icarus},
  volume={311},
  pages={1--22},
  year={2018},
  publisher={Elsevier}
}

@article{giles2024,
  title={Seasonal and longitudinal variability in {Io}’s {SO$_2$} atmosphere from 22 years of {IRTF/TEXES} observations},
  author={Giles, Rohini S and Spencer, John R and Tsang, Constantine CC and Greathouse, Thomas K and Lellouch, Emmanuel and L{\'o}pez-Valverde, Miguel A},
  journal={Icarus},
  volume={418},
  pages={116151},
  year={2024},
  publisher={Elsevier}
}

@article{hamilton2025,
  title={Comparing {NASA} {D}iscovery and {N}ew {F}rontiers {C}lass {M}ission {C}oncepts for the {I}o {V}olcano {O}bserver},
  author = {Hamilton, Christopher W. and McEwen, Alfred S. and Keszthelyi, Laszlo and Carter, Lynn M. and Davies, Ashley G. and de Kleer, Katherine and Jessup, Kandis Lea and Jia, Xianzhe and Keane, James T. and Mandt, Kathleen and Nimmo, Francis and Paranicas, Chris and Park, Ryan S. and Perry, Jason E. and Pommier, Anne and Radebaugh, Jani and Sutton, Sarah S. and Vorburger, Audrey and Wurz, Peter and Borlina, Cauê and Haapala, Amanda F. and DellaGiustina, Daniella N. and Denevi, Brett W. and Hörst, Sarah M. and Kempf, Sascha and Khurana, Krishan K. and Likar, Justin J. and Masters, Adam and Mousis, Olivier and Polit, Anjani T. and Bhushan, Aditya and Bland, Michael and Matsuyama, Isamu and Spencer, John},
  journal={The Planetary Science Journal},
  volume={6},
  number={6},
  pages={134},
  year={2025},
  publisher={IOP Publishing}
}

@article{hoey2016,
	title = {{DSMC} simulation of {Io}’s unsteady {Tvashtar} plume},
	volume = {1786},
	issn = {0094-243X},
	url = {https://doi.org/10.1063/1.4967663},
	doi = {10.1063/1.4967663},
	number = {1},
	journal = {AIP Conference Proceedings},
	author = {Hoey, W. A. and Ackley, P. C. and Trafton, L. M. and Goldstein, D. B. and Varghese, P. L.},
	month = nov,
	year = {2016},
	pages = {160006},
}

@article{hoey2021,
  title={Variations in the canopy shock structures of massive extraterrestrial plumes: Parametric {DSMC} simulation of 2007 {Tvashtar} observations},
  author={Hoey, William A and Trafton, Laurence M and Ackley, Peter C and Goldstein, David B and Varghese, Philip L},
  journal={Icarus},
  volume={363},
  pages={114431},
  year={2021},
  publisher={Elsevier}
}

@article{jessup2004,
  title={The atmospheric signature of {Io}'s {Prometheus} plume and anti-jovian hemisphere: evidence for a sublimation atmosphere},
  author={Jessup, Kandis Lea and Spencer, John R and Ballester, Gilda E and Howell, Robert R and Roesler, Fred and Vigel, Miquela and Yelle, Roger},
  journal={Icarus},
  volume={169},
  number={1},
  pages={197--215},
  year={2004},
  publisher={Elsevier}
}

@article{jessup2007,
  title={Sulfur volcanism on {Io}},
  author={Jessup, Kandis Lea and Spencer, John and Yelle, Roger},
  journal={Icarus},
  volume={192},
  number={1},
  pages={24--40},
  year={2007},
  publisher={Elsevier}
}

@article{keane2022,
  title={Perspective: the future exploration of {Io}},
  author={Keane, James Tuttle and de Kleer, Katherine and Spencer, John},
  journal={Elements: An International Magazine of Mineralogy, Geochemistry, and Petrology},
  volume={18},
  number={6},
  pages={399--404},
  year={2022},
  publisher={Mineralogical Society of America}
}

@article{kieffer1982,
  title={Dynamics and thermodynamics of volcanic eruptions-{Implications} for the plumes on {Io}},
  author={Kieffer, SW},
  journal={Satellites of {Jupiter}},
  pages={647--723},
  year={1982}
}

@phdthesis{klaiber2024,
  title={Three dimensional {DSMC} modelling of the dynamics of {I}o’s atmosphere: Building a simulation setup for parameter studies of the effects and the interaction of different processes in {I}o’s atmosphere},
  author={Klaiber, Lea Meera},
  year={2024},
  school={Universit{\"a}t Bern},
  doi={10.48549/4991}
}

@article{kumar1982,
  title={The atmospheres of {Io} and other satellites},
  author={Kumar, S and Hunten, DM},
  journal={Satellites of {Jupiter}},
  pages={782--806},
  year={1982}
}

@article{lellouch1990,
  title={{Io}'s atmosphere from microwave detection of {SO$_2$}},
  author={Lellouch, Emmanuel and Belton, Michael and de Pater, Imke and Gulkis, Samuel and Encrenaz, Th{\'e}r{\`e}se},
  journal={Nature},
  volume={346},
  number={6285},
  pages={639--641},
  year={1990},
  publisher={Nature Publishing Group UK London}
}

@article{lellouch1992,
  title={The structure, stability, and global distribution of {Io}'s atmosphere},
  author={Lellouch, Emmanuel and Belton, Michael and De Pater, Imke and Paubert, Gabriel and Gulkis, Samuel and Encrenaz, Th{\'e}r{\`e}se},
  journal={Icarus},
  volume={98},
  number={2},
  pages={271--295},
  year={1992},
  publisher={Elsevier}
}

@article{lellouch1996,
  title={Detection of sulfur monoxide in {Io}'s atmosphere},
  author={Lellouch, E and Strobel, DF and Belton, MJS and Summers, ME and Paubert, G and Moreno, R},
  journal={The Astrophysical Journal},
  volume={459},
  number={2},
  pages={L107},
  year={1996},
  publisher={IOP Publishing}
}

@article{lellouch2003,
  title={Volcanically emitted sodium chloride as a source for {Io}'s neutral clouds and plasma torus},
  author={Lellouch, Emmanuel and Paubert, Gabriel and Moses, Julianne I and Schneider, Nicholas M and Strobel, DF},
  journal={Nature},
  volume={421},
  number={6918},
  pages={45--47},
  year={2003},
  publisher={Nature Publishing Group UK London}
}

@article{lellouch2015,
  title={Detection and characterization of {Io}’s atmosphere from high-resolution 4-$\mu$m spectroscopy},
  author={Lellouch, E and Ali-Dib, M and Jessup, K-L and Smette, A and K{\"a}ufl, H-U and Marchis, F},
  journal={Icarus},
  volume={253},
  pages={99--114},
  year={2015},
  publisher={Elsevier}
}

@article{marschall2019,
  title={A comparison of multiple {R}osetta data sets and 3D model calculations of {67P/Churyumov-Gerasimenko} coma around equinox (May 2015)},
  author={Marschall, Raphael and Rezac, Ladislav and Kappel, David and Su, Chin-Chia and Gerig, S-B and Rubin, Martin and Pinz{\'o}n-Rodr{\'\i}guez, O and Marshall, David and Liao, Ying and Herny, Cl{\'e}mence and others},
  journal={Icarus},
  volume={328},
  pages={104--126},
  year={2019},
  publisher={Elsevier}
}

@article{mcgrath2000,
  title={Spatially resolved spectroscopy of {Io}'s {Pele} plume and {SO$_2$} atmosphere},
  author={McGrath, Melissa A and Belton, Michael JS and Spencer, John R and Sartoretti, Paola},
  journal={Icarus},
  volume={146},
  number={2},
  pages={476--493},
  year={2000},
  publisher={Elsevier}
}

@article{mokhtari2025,
  title={The role of the hot porous layer in the gas flow in the inner coma},
  author={Mokhtari, O and Skorov, Yu V and Rezac, L and Marschall, R and Belousov, D and Pinz{\'o}n-Rodr{\'\i}guez, O and K{\"u}ppers, M and Thomas, N},
  journal={Astronomy \& Astrophysics},
  volume={693},
  pages={A57},
  year={2025},
  publisher={EDP Sciences}
}

@article{moore2009,
  title={1-{D} {DSMC} simulation of {Io}'s atmospheric collapse and reformation during and after eclipse},
  author={Moore, CH and Goldstein, DB and Varghese, PL and Trafton, LM and Stewart, B},
  journal={Icarus},
  volume={201},
  number={2},
  pages={585--597},
  year={2009},
  publisher={Elsevier}
}

@inproceedings{moore2012,
  title={{DSMC} simulations of the plasma bombardment on {Io}'s sublimated and sputtered atmosphere},
  author={Moore, Chris and Walker, Andrew and Goldstein, David and Varghese, Philip and Trafton, Laurence and Parsons, Neal and Levin, Deborah},
  booktitle={50th {AIAA} {A}erospace {S}ciences {M}eeting including the {N}ew {H}orizons {F}orum and {A}erospace {E}xposition},
  doi = {10.2514/6.2012-560},
  pages={560},
  year={2012}
}

@article{moses2002,
	title = {Photochemistry of a {Volcanically} {Driven} {Atmosphere} on {Io}: {Sulfur} and {Oxygen} {Species} from a {Pele}-{Type} {Eruption}},
	volume = {156},
	issn = {0019-1035},
	url = {https://www.sciencedirect.com/science/article/pii/S0019103501967582},
	doi = {https://doi.org/10.1006/icar.2001.6758},
	number = {1},
	journal = {Icarus},
	author = {Moses, Julianne I. and Zolotov, Mikhail Yu and Fegley, Bruce},
	year = {2002},
	pages = {76--106},
}

@article{moullet2008,
  title={First disk-resolved millimeter observations of {Io}'s surface and {SO$_2$} atmosphere},
  author={Moullet, Arielle and Lellouch, Emmanuel and Moreno, Rapha{\"e}l and Gurwell, Mark A and Moore, C},
  journal={Astronomy \& Astrophysics},
  volume={482},
  number={1},
  pages={279--292},
  year={2008},
  publisher={EDP Sciences}
}

@article{moullet2010,
  title={Simultaneous mapping of {SO$_2$}, {SO}, {NaCl} in {Io}’s atmosphere with the {Submillimeter Array}},
  author={Moullet, Arielle and Gurwell, Mark A and Lellouch, Emmanuel and Moreno, Rapha{\"e}l},
  journal={Icarus},
  volume={208},
  number={1},
  pages={353--365},
  year={2010},
  publisher={Elsevier}
}

@article{moullet2013,
  title={Exploring {Io}'s atmospheric composition with {APEX: First} measurement of $^{34}${SO}${_2}$ and tentative detection of {KCl}},
  author={Moullet, A. and Lellouch, E and Moreno, R and Gurwell, M and Black, John H and Butler, B},
  journal={The Astrophysical Journal},
  volume={776},
  number={1},
  pages={32},
  year={2013},
  publisher={IOP Publishing}
}

@article{pinzon_rodriguez2021,
  title={The effect of thermal conductivity on the outgassing and local gas dynamics from cometary nuclei},
  author={Pinz{\'o}n-Rodr{\'\i}guez, O and Marschall, R and Gerig, S-B and Herny, C and Wu, JS and Thomas, N},
  journal={Astronomy \& Astrophysics},
  volume={655},
  pages={A20},
  year={2021},
  publisher={EDP Sciences}
}

@article{rathbun2004,
  title={Mapping of {Io}'s thermal radiation by the {Galileo} photopolarimeter--radiometer ({PPR}) instrument},
  author={Rathbun, JA and Spencer, JR and Tamppari, LK and Martin, TZ and Barnard, L and Travis, LD},
  journal={Icarus},
  volume={169},
  number={1},
  pages={127--139},
  year={2004},
  publisher={Elsevier}
}

@article{redwing2022,
  title={{NaCl} and {KCl} in {Io’s} atmosphere},
  author={Redwing, Erin and de Pater, Imke and Luszcz-Cook, Statia and de Kleer, Katherine and Moullet, Arielle and Rojo, Patricio M},
  journal={The Planetary Science Journal},
  volume={3},
  number={10},
  pages={238},
  year={2022},
  publisher={IOP Publishing}
}

@article{roth2020,
  title={An attempt to detect transient changes in {Io}’s {SO$_2$} and {NaCl} atmosphere},
  author={Roth, Lorenz and Boissier, Jeremie and Moullet, Arielle and S{\'a}nchez-Monge, {\'A}lvaro and de Kleer, Katherine and Yoneda, Mizuki and Hikida, Reina and Kita, Hajime and Tsuchiya, Fuminori and Bl{\"o}cker, Aljona and others},
  journal={Icarus},
  volume={350},
  pages={113925},
  year={2020},
  publisher={Elsevier}
}

@article{saur2004,
	title = {Relative contributions of sublimation and volcanoes to {Io}'s atmosphere inferred from its plasma interaction during solar eclipse},
	volume = {171},
	copyright = {https://www.elsevier.com/tdm/userlicense/1.0/},
	issn = {00191035},
	url = {https://linkinghub.elsevier.com/retrieve/pii/S0019103504001630},
	doi = {10.1016/j.icarus.2004.05.010},
	language = {en},
	number = {2},
	journal = {Icarus},
	author = {Saur, Joachim and Strobel, Darrell F.},
	month = oct,
	year = {2004},
	pages = {411--420},
}

@article{schenk2001,
  title={The mountains of {Io}: Global and geological perspectives from {Voyager} and {Galileo}},
  author={Schenk, Paul and Hargitai, Henrik and Wilson, Ronda and McEwen, Alfred and Thomas, Peter},
  journal={Journal of Geophysical Research: Planets},
  volume={106},
  number={E12},
  pages={33201--33222},
  year={2001},
  publisher={Wiley Online Library}
}

@misc{schlarmann_2025,
  author       = {Schlarmann, Leander},
  title        = {Interactions of sublimated frost with volcanic
                   plumes: {M}odelling {I}o's {SO2} atmosphere using the {DSMC} method
                  },
  month        = dec,
  year         = 2025,
  publisher    = {Zenodo},
  version      = {v1},
  doi          = {10.5281/zenodo.17790309},
  url          = {https://doi.org/10.5281/zenodo.17790309},
}

@article{schmitt2003,
  title={Possible identification of local deposits of {Cl$_2$SO$_2$} on {Io} from {NIMS/Galileo} spectra},
  author={Schmitt, B and Rodriguez, S},
  journal={Journal of Geophysical Research: Planets},
  volume={108},
  number={E9},
  year={2003},
  publisher={Wiley Online Library}
}

@article{schorghofer2024,
  title={Sublimation pressures of common volatiles at low temperature and maps of supervolatile cold traps on the {M}oon},
  author={Sch{\"o}rghofer, Norbert and Williams, Jean-Pierre},
  journal={Icarus},
  volume={416},
  pages={116086},
  year={2024},
  publisher={Elsevier}
}

@article{spencer1989,
  title={Systematic biases in radiometric diameter determinations},
  author={Spencer, John R and Lebofsky, Larry A and Sykes, Mark V},
  journal={Icarus},
  volume={78},
  number={2},
  pages={337--354},
  year={1989},
  publisher={Elsevier}
}

@article{spencer1992,
  title={The influence of thermal inertia on temperatures and frost stability on Triton},
  author={Spencer, John R and Moore, Jeffrey M},
  journal={Icarus},
  volume={99},
  number={2},
  pages={261--272},
  year={1992},
  publisher={Elsevier}
}

@article{spencer1997,
  title={The {Pele} plume ({Io}): Observations with the {Hubble} {Space} {Telescope}},
  author={Spencer, John R and Sartoretti, Paola and Ballester, Gilda E and McEwen, Alfred S and Clarke, John T and McGrath, Melissa A},
  journal={Geophysical research letters},
  volume={24},
  number={20},
  pages={2471--2474},
  year={1997},
  publisher={Wiley Online Library}
}

@article{spencer2000,
  title={Discovery of gaseous {S$_2$} in {Io}'s {Pele} plume},
  author={Spencer, John R and Jessup, Kandis Lea and McGrath, Melissa A and Ballester, Gilda E and Yelle, Roger},
  journal={Science},
  volume={288},
  number={5469},
  pages={1208--1210},
  year={2000},
  publisher={American Association for the Advancement of Science}
}

@article{spencer2005,
  title={Mid-infrared detection of large longitudinal asymmetries in {Io}'s {SO$_2$} atmosphere},
  author={Spencer, John R and Lellouch, Emmanuel and Richter, Matthew J and L{\'o}pez-Valverde, Miguel A and Jessup, Kandis Lea and Greathouse, Thomas K and Flaud, Jean-Marie},
  journal={Icarus},
  volume={176},
  number={2},
  pages={283--304},
  year={2005},
  publisher={Elsevier}
}

@article{stull1947,
    author = {Stull, Daniel R.},
    title = {Inorganic Compounds},
    journal = {Industrial \& Engineering Chemistry},
    volume = {39},
    number = {4},
    pages = {540-550},
    year = {1947},
    doi = {10.1021/ie50448a023},
    URL = {https://doi.org/10.1021/ie50448a023},    
    eprint = {https://doi.org/10.1021/ie50448a023}
}

@article{strobel1994,
	title = {On the {Vertical} {Thermal} {Structure} of {Io}'s {Atmosphere}},
	volume = {111},
	copyright = {https://www.elsevier.com/tdm/userlicense/1.0/},
	issn = {00191035},
	url = {https://linkinghub.elsevier.com/retrieve/pii/S0019103584711304},
	doi = {10.1006/icar.1994.1130},
	language = {en},
	number = {1},
	journal = {Icarus},
	author = {Strobel, Darrell F. and Zhu, Xun and Summers, Michael E.},
	month = sep,
	year = {1994},
	pages = {18--30},
}

@article{strobel2001,
  title={The atmosphere of {Io}: Abundances and sources of sulfur dioxide and atomic hydrogen},
  author={Strobel, Darrell F and Wolven, Brian C},
  journal={Astrophysics and Space Science},
  volume={277},
  pages={271--287},
  year={2001},
  publisher={Springer}
}

@article{sieveka1985,
	title = {Nonisotropic coronal atmosphere on {Io}},
	volume = {90},
	copyright = {http://onlinelibrary.wiley.com/termsAndConditions\#vor},
	issn = {0148-0227},
	url = {https://agupubs.onlinelibrary.wiley.com/doi/10.1029/JA090iA06p05327},
	doi = {10.1029/JA090iA06p05327},
	language = {en},
	number = {A6},
	journal = {Journal of Geophysical Research: Space Physics},
	author = {Sieveka, E. M. and Johnson, R. E.},
	month = jun,
	year = {1985},
	pages = {5327--5331},
}

@article{sinton1988,
  title={Infrared observations of eclipses of {Io}, its thermophysical parameters, and the thermal radiation of the {Loki} volcano and environs},
  author={Sinton, William M and Kaminski, Charles},
  journal={Icarus},
  volume={75},
  number={2},
  pages={207--232},
  year={1988},
  publisher={Elsevier}
}

@article{simonelli1984,
  title={Voyager disk-integrated photometry of {Io}},
  author={Simonelli, Damon P and Veverka, Joseph},
  journal={Icarus},
  volume={59},
  number={3},
  pages={406--425},
  year={1984},
  publisher={Elsevier}
}

@article{simonelli1988,
  title={Bolometric albedos and diurnal temperatures of the brightest regions on {Io}},
  author={Simonelli, Damon P and Veverka, Joseph},
  journal={Icarus},
  volume={74},
  number={2},
  pages={240--261},
  year={1988},
  publisher={Elsevier}
}

@article{simonelli2001,
  title={Regolith variations on {Io}: Implications for bolometric albedos},
  author={Simonelli, Damon P and Dodd, Christopher and Veverka, Joseph},
  journal={Journal of Geophysical Research: Planets},
  volume={106},
  number={E12},
  pages={33241--33252},
  year={2001},
  publisher={Wiley Online Library}
}

@article{strom1982,
  title={Volcanic eruption plumes on {Io}},
  author={Strom, Robert G and Schneider, NM},
  journal={Satellites of {Jupiter}},
  pages={598--633},
  year={1982}
}

@article{su2013,
  title={Parallel {D}irect {S}imulation {M}onte {C}arlo ({DSMC}) methods for modeling rarefied gas dynamics},
  author={Su, CC},
  journal={Taiwan: National Chiao Tung University},
  year={2013}
}

@article{thomas1987,
  title={Condensation and sublimation on {Io}--{I}},
  author={Thomas, Nicolas},
  journal={Monthly Notices of the Royal Astronomical Society},
  volume={226},
  number={1},
  pages={195--207},
  year={1987},
  publisher={The Royal Astronomical Society}
}

@article{thomas2004,
  title={The {Io} neutral clouds and plasma torus},
  author={Thomas, N and Bagenal, F and Hill, TW and Wilson, JK},
  journal={Jupiter. The {P}lanet, {S}atellites and {M}agnetosphere},
  volume={1},
  pages={561--591},
  year={2004}
}

@article{trafton1996,
  title={The Gaseous Sulfur Dioxide Abundance over {Io}'s Leading and Trailing Hemispheres: {HST} Spectra of {Io}'s {C}$^1${B}$_2$--{X}$^1${A}$_1$ Band of {SO$_2$} near 2100 {A}ngstrom},
  author={Trafton, LM and Caldwell, JJ and Barnet, C and Cunningham, CC},
  journal={Astrophysical Journal v. 456, p. 384},
  volume={456},
  pages={384},
  year={1996}
}

@article{trumbo2022,
  title={Spectroscopic Mapping of {Io}’s Surface with {HST/STIS}: {SO$_2$} Frost, Sulfur Allotropes, and Large-scale Compositional Patterns},
  author={Trumbo, Samantha K and Davis, M Ryleigh and Cassese, Benjamin and Brown, Michael E},
  journal={The Planetary Science Journal},
  volume={3},
  number={12},
  pages={272},
  year={2022},
  publisher={IOP Publishing}
}

@article{tsang2012,
  title={Io’s atmosphere: {Constraints} on sublimation support from density variations on seasonal timescales using {NASA} {IRTF/TEXES} observations from 2001 to 2010},
  author={Tsang, Constantine CC and Spencer, John R and Lellouch, Emmanuel and L{\'o}pez-Valverde, Miguel A and Richter, Matthew J and Greathouse, Thomas K},
  journal={Icarus},
  volume={217},
  number={1},
  pages={277--296},
  year={2012},
  publisher={Elsevier}
}

@article{tsang2013,
  title={Io’s contracting atmosphere post 2011 perihelion: Further evidence for partial sublimation support on the anti-{Jupiter} hemisphere},
  author={Tsang, Constantine CC and Spencer, John R and Lellouch, Emmanuel and L{\'o}pez-Valverde, Miguel A and Richter, Matthew J and Greathouse, Thomas K and Roe, Henry},
  journal={Icarus},
  volume={226},
  number={1},
  pages={1177--1181},
  year={2013},
  publisher={Elsevier}
}

@article{tsang2015,
  title={Non-detection of post-eclipse changes in {Io}’s {Jupiter}-facing atmosphere: Evidence for volcanic support?},
  author={Tsang, Constantine CC and Spencer, John R and Jessup, Kandis Lea},
  journal={Icarus},
  volume={248},
  pages={243--253},
  year={2015},
  publisher={Elsevier}
}

@article{tsang2016,
  title={The collapse of {Io}'s primary atmosphere in {Jupiter} eclipse},
  author={Tsang, Constantine CC and Spencer, John R and Lellouch, Emmanuel and Lopez-Valverde, Miguel A and Richter, Matthew J},
  journal={Journal of Geophysical Research: Planets},
  volume={121},
  number={8},
  pages={1400--1410},
  year={2016},
  publisher={Wiley Online Library}
}

@article{wagman1979,
  title={Sublimation pressure and enthalpy of {SO$_2$}},
  author={Wagman, DD},
  journal={Chem. Thermodynamics Data Center Natl. Bureau Standards, Washington, DC},
  year={1979}
}

@article{walker2010,
  title={A comprehensive numerical simulation of {Io}’s sublimation-driven atmosphere},
  author={Walker, Andrew C and Gratiy, Sergey L and Goldstein, David B and Moore, Chris H and Varghese, Philip L and Trafton, Laurence M and Levin, Deborah A and Stewart, B{\'e}n{\'e}dicte},
  journal={Icarus},
  volume={207},
  number={1},
  pages={409--432},
  year={2010},
  publisher={Elsevier}
}

@article{walker2012,
	title = {A parametric study of {Io}’s thermophysical surface properties and subsequent numerical atmospheric simulations based on the best fit parameters},
	volume = {220},
	copyright = {https://www.elsevier.com/tdm/userlicense/1.0/},
	issn = {00191035},
	url = {https://linkinghub.elsevier.com/retrieve/pii/S0019103512001741},
	doi = {10.1016/j.icarus.2012.05.001},
	language = {en},
	number = {1},
	journal = {Icarus},
	author = {Walker, Andrew C. and Moore, Chris H. and Goldstein, David B. and Varghese, Philip L. and Trafton, Laurence M.},
	month = jul,
	year = {2012},
	pages = {225--253},
}

@article{wolven2001,
  title={Emission profiles of neutral oxygen and sulfur in Io's exospheric corona},
  author={Wolven, BC and Moos, HW and Retherford, KD and Feldman, PD and Strobel, DF and Smyth, WH and Roesler, FL},
  journal={Journal of Geophysical Research: Space Physics},
  volume={106},
  number={A11},
  pages={26155--26182},
  year={2001},
  publisher={Wiley Online Library}
}

@article{wong1995,
	title = {The {Effect} of {Plasma} {Heating} on {Sublimation}-{Driven} {Flow} in {Io}'s {Atmosphere}},
	volume = {115},
	copyright = {https://www.elsevier.com/tdm/userlicense/1.0/},
	issn = {00191035},
	url = {https://linkinghub.elsevier.com/retrieve/pii/S0019103585710822},
	doi = {10.1006/icar.1995.1082},
	language = {en},
	number = {1},
	journal = {Icarus},
	author = {Wong, Mau C. and Johnson, Robert E.},
	month = may,
	year = {1995},
	pages = {109--118},
}

@article{wong1996_pc,
	title = {A three‐dimensional azimuthally symmetric model atmosphere for {Io}: 1. {Photochemistry} and the accumulation of a nightside atmosphere},
	volume = {101},
	copyright = {http://onlinelibrary.wiley.com/termsAndConditions\#vor},
	issn = {0148-0227},
	shorttitle = {A three‐dimensional azimuthally symmetric model atmosphere for {Io}},
	url = {https://agupubs.onlinelibrary.wiley.com/doi/10.1029/96JE02510},
	doi = {10.1029/96JE02510},
	language = {en},
	number = {E10},
	journal = {Journal of Geophysical Research: Planets},
	author = {Wong, Mau C. and Johnson, Robert E.},
	month = oct,
	year = {1996},
	pages = {23243--23254},
}

@article{wong1996_plasma,
	title = {A three‐dimensional azimuthally symmetric model atmosphere for {Io}: 2. {Plasma} effect on the surface},
	volume = {101},
	copyright = {http://onlinelibrary.wiley.com/termsAndConditions\#vor},
	issn = {0148-0227},
	shorttitle = {A three‐dimensional azimuthally symmetric model atmosphere for {Io}},
	url = {https://agupubs.onlinelibrary.wiley.com/doi/10.1029/96JE02509},
	doi = {10.1029/96JE02509},
	language = {en},
	number = {E10},
	urldate = {2025-05-30},
	journal = {Journal of Geophysical Research: Planets},
	author = {Wong, Mau C. and Johnson, Robert E.},
	month = oct,
	year = {1996},
	pages = {23255--23259},
}

@article{wu2003,
  title={Parallel three-dimensional direct simulation {M}onte {C}arlo method and its applications},
  author={Wu, Jong-Shinn and Lian, Y-Y},
  journal={Computers \& Fluids},
  volume={32},
  number={8},
  pages={1133--1160},
  year={2003},
  publisher={Elsevier}
}

@article{wu2004,
  title={Parallel three-dimensional {DSMC} method using mesh refinement and variable time-step scheme},
  author={Wu, Jong-Shinn and Tseng, Kun-Chang and Wu, Fu-Yuan},
  journal={Computer Physics Communications},
  volume={162},
  number={3},
  pages={166--187},
  year={2004},
  publisher={Elsevier}
}

@article{wu2005,
  title={Parallel {DSMC} method using dynamic domain decomposition},
  author={Wu, Jong-Shinn and Tseng, K-C},
  journal={International Journal for numerical methods in Engineering},
  volume={63},
  number={1},
  pages={37--76},
  year={2005},
  publisher={Wiley Online Library}
}

@article{zhang2003,
  title={Simulation of gas dynamics and radiation in volcanic plumes on {Io}},
  author={Zhang, J and Goldstein, DB and Varghese, PL and Gimelshein, NE and Gimelshein, SF and Levin, DA},
  journal={Icarus},
  volume={163},
  number={1},
  pages={182--197},
  year={2003},
  publisher={Elsevier}
}

@article{zhang2004,
  title={Numerical modeling of ionian volcanic plumes with entrained particulates},
  author={Zhang, J and Goldstein, DB and Varghese, PL and Trafton, L and Moore, C and Miki, K},
  journal={Icarus},
  volume={172},
  number={2},
  pages={479--502},
  year={2004},
  publisher={Elsevier}
}

@article{zolotov1998,
  title={Volcanic production of sulfur monoxide ({SO}) on {I}o},
  author={Zolotov, Mikhail Yu and Fegley Jr, Bruce},
  journal={Icarus},
  volume={132},
  number={2},
  pages={431--434},
  year={1998},
  publisher={Elsevier}
}

%
%
%
%
%

\end{document}